\documentclass[fleqn,usenatbib]{mnras}

\usepackage{newtxtext,newtxmath}

\usepackage[T1]{fontenc}

\DeclareRobustCommand{\VAN}[3]{#2}
\let\VANthebibliography\thebibliography
\def\thebibliography{\DeclareRobustCommand{\VAN}[3]{##3}\VANthebibliography}

\usepackage{graphicx}	
\usepackage{amsmath}	

\newcommand\mbh{M_{\bullet}}

\newcommand\teff{T_{\rm eff}}
\newcommand\rin{r_{\rm in}}
\newcommand\sigB{\sigma_{\rm B}}

\newcommand\re{R_{\rm E}}
\newcommand\hp{h}

\newcommand\msun{M_{\odot}}

\newcommand\abbh{a_{\rm BBH}}

\newcommand\be{\begin{equation}}
\newcommand\ee{\end{equation}}
\newcommand\bea{\begin{eqnarray}}
\newcommand\eea{\end{eqnarray}}

\title[An equal mass ratio SMBBHs in Q J0158-4325]{An equal mass ratio supermassive binary black holes in Q J0158-4325 with periodic microlensing signature?}

\author[Yan et al.]{
Changshuo Yan$^{1,2}$\thanks{E-mail: yancs@nao.cas.cn},
Youjun Lu$^{2,1}$\thanks{E-mail: luyj@nao.cas.cn},
Junqiang Ge$^{1}$,
and Erlin Qiao$^{1,2}$
\\
$^{1}$National Astronomical Observatories, Chinese Academy of Sciences, 20A Datun Road, Beijing 100101, China\\
$^{2}$School of Astronomy and Space Science, University of Chinese Academy of Sciences, 19A Yuquan Road, Beijing 100049, China
}

\date{Accepted XXX. Received YYY; in original form ZZZ}

\pubyear{\the\year{}}

\begin{document}
\label{firstpage}
\pagerange{\pageref{firstpage}--\pageref{lastpage}}
\maketitle

\begin{abstract}
This study aims to test whether a supermassive binary black hole (SMBBH) system with a triple-disk accretion structure can explain the observed $\sim$173-day periodic microlensing variations and spectral energy distribution (SED) of the gravitationally lensed quasar Q J0158-4325. We construct a triple-disk model for the SMBBH system, incorporating realistic accretion disk structures, orbital motion, and microlensing effects. The model is used to simulate optical and X-ray microlensing light curves and SEDs, which are compared with long-term optical monitoring, X-ray observations, and UV-optical spectra from HST and XSHOOTER. Bayesian analysis and MCMC fitting are applied to constrain model parameters. The model successfully reproduces the periodic microlensing variations. Combined light curve and SED fitting favor a high mass ratio ($q>0.5$) SMBBH system with total mass $\sim 10^{9.5}M_\odot$, and nearly equal-mass binaries ($q\sim1$) provides the best agreement with both the optical/UV spectrum and the microlensing signal. This model predicts larger X-ray microlensing amplitudes than in the optical, but, the available X-ray observations lack the precision needed to place strong constraints. We emphasize the need for future high-cadence monitoring to resolve remaining uncertainties. This study demonstrates the effectiveness of combining multi-wavelength microlensing signatures with spectral modeling to provide robust constraints on SMBBH systems, with the developed framework applicable to other lensed quasars for identifying and characterizing candidate SMBBHs.
\end{abstract}




\section{Introduction}
\label{sec:intro}

The $\Lambda$CDM cosmological model predicts hierarchical galaxy formation through mergers, inevitably leading to the formation of supermassive binary black holes (SMBBHs) if each progenitor galaxy contains a central supermassive black hole (SMBH) \citep[e.g.,][]{BBR80,Yu02,MM05,2020ApJ...897...86C}. In gas-rich mergers, these SMBBHs become embedded in circumbinary disks with central cavities cleared by the secondary SMBH's gravitational influence \citep{AL94,Ivanov99,Escala05}. Within these cavities, each SMBH may maintain its own mini-disk, fed by gas streams from the circumbinary disk \citep{Hayasaki07,Hayasaki08,Dotti07,Cuadra09,DOrazio12,Farris14,Dittmann2023,Siwek23}. Such a triple-disk system may produce a unique SED characterized by UV/optical flux deficit due to the truncated disk geometry \citep[e.g.,][]{GM12,Sesana12,Hayasaki13,Roedig14,Yan2014,Yan2015,Farris15,Zheng2016,Krauth23}, providing distinct observational signatures of such systems.

Microlensing in gravitationally lensed quasars has emerged as a powerful diagnostic tool for studying accretion disk structure at microarcsecond scales \citep{WPS90, WMS95, Wambsganss06}. The Einstein radii of stars in lensing galaxies ($\sim10^{16}$ cm) conveniently match the expected size scales of quasar accretion disks, enabling detailed studies of disk temperature profiles and radial structure through multi-wavelength monitoring \citep{Morgan12,Chartas2017,Guerras2017,Cornachione2020}. As first proposed by \citet{Yan2015}, the orbital motion of SMBBH-triple disk systems should imprint characteristic periodic signatures in microlensing light curves and deviation in size-wavelength relation, offering a novel method to detect these compact binaries. The periodicity of these fluctuations is determined by the mass ratio of the binary black holes. Systems with nearly equal masses exhibit variations at half the orbital period, whereas those with a low mass ratio the period corresponds to the orbital period influenced by the secondary mini-disk. 

Recent observations of the lensed quasar Q J0158-4325 revealed a $\sim$173-day periodic oscillations in its $15$-year optical light curve \citep{Millon2022}. While the simplified SMBBH model in \citet{Millon2022} - featuring a primary accretion disk with an orbiting hotspot representing the secondary - could reproduce the observed periodicity, several key aspects require more rigorous examination. A robust SMBBH interpretation must simultaneously account for the system's dynamical consistency (where the observed orbital period matches predictions from the SMBBH mass and separation), produce multi-wavelength microlensing signatures consistent with triple-disk geometry, and demonstrate SED features that remain viable after considering dust extinction and Fe II contamination.

In this work, we aim to revisit the SMBBH model for the microlensing light curve of Q J0158-4325 by considering relatively more realistic SMBBH-triple disk structure of the system. We investigate whether the SMBBH-triple disk system can provide a good fit to the observed light curve by incorporating detailed physical parameters and mechanisms and check whether such a model produce an SED consistent with the observations. The organization of this paper is as follows. In Section~\ref{sec:J0158}, we list the observation data of Q J0158-4325. In Section~\ref{sec:model}, we briefly describe the simple model for the active SMBBH-triple disk system to calculate the surface brightness distribution of the system in the optical bands and the spectral energy density distribution. In Section~\ref{sec:result}, we reconstruct the observed microlensing light curves and SED of Q J0158-4325 and provide predictions for the X-ray microlensing light curves. We discuss some limitations of the model in Section~\ref{sec:dis}. Our main conclusions are summarized in Section~\ref{sec:con}.

\section{Observational data of Q J0158-4325}

Q J0158-4325 is a gravitationally lensed quasar at $ z_{\rm obs} = 1.29 $, with two images (A and B) separated by $ 1\farcs3 $. The system exhibits periodic oscillations in its $15$-year optical light curves with a period of $172.6\pm0.9$ days, interpreted by \citet{Millon2022} as potential microlensing signatures from an SMBBH system. Their model suggested a primary SMBH accreting via a standard thin disk, with a secondary SMBH generating periodic hotspots through orbital motion.

The lensing configuration is characterized by convergence and shear values of $(\kappa, \gamma) = (0.72, 1.03)$ for image B and $(0.23, 0.39)$ for image A \citep{Morgan08}. Using $\mu = [(1-\kappa)^2 - \gamma^2]^{-1}$, we derive macrolensing magnifications $\mu_{\rm macro,B} \approx -1.02$ and $\mu_{\rm macro,A} \approx 2.23$, corresponding to a magnitude difference $\Delta m_0 = 0.87$. Following \citet{Millon2022}, we isolate microlensing variations in image B by subtracting intrinsic variability traced through image A's light curve.

\begin{figure}
\centering
\includegraphics[width=0.23\textwidth,angle=-90]{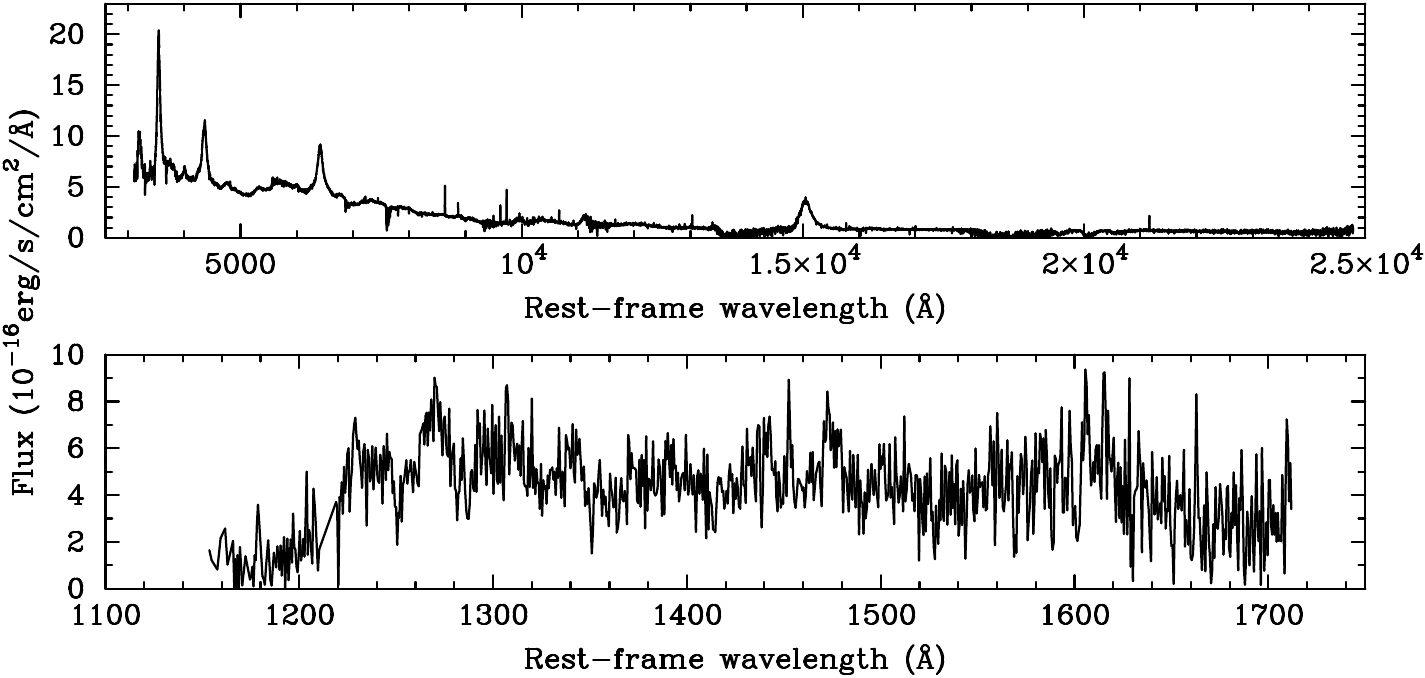}
\caption{The XSHOOTER spectrum (top panel) and XSHOOTER spectrum (bottom panel) of Q J0158-4325 in the observer's frame.
}
\label{fig:observation}
\end{figure}

Q J0158-4325 was observed using XSHOOTER in 2019 August 22, resulting in a set of six spectra. Three of these spectra were captured at the coordinates $\alpha_{\rm J2000}=01:58:41.42$ and $\delta_{\rm J2000}=-43:25:02.1$, while the other three were taken at $\alpha_{\rm J2000}=01:58:41.50$ and $\delta_{\rm J2000}=-43:25:01.0$. The positions of the images are as follows: image A is located at $\alpha_{\rm J2000}=01:58:41.43$ and $\delta_{\rm J2000}=-43:25:03.4$, and image B is at $\alpha_{\rm J2000}=01:58:41.32$ and $\delta_{\rm J2000}=-43:25:03.8$ \citep{Morgan1999}. Consequently, it is likely that these six spectra are primarily derived from image A. The exposure times for the spectra range from 1180 to 2400 seconds, with a spectral resolution of $0.06~\AA{\rm pix}^{-1}$ for wavelengths between $994$ nm and $2479$ nm, and $0.02~\AA{\rm pix}^{-1}$ for the ranges of $298.9$ nm to $556$ nm and $533.7$ nm to $1020$ nm. We re-bin the spectra to a resolution of $1$\AA, which is presented in the top panel of Figure~\ref{fig:observation}. 
The spectrum was then corrected for Galactic extinction using an $E(B-V)$ value of 0.022 mag, as reported by \cite{Schlafly11} in the NASA/IPAC Extragalactic Database (NED), along with the reddening curve from \cite{Fitzpatrick99}. We then applied corrections for macrolensing magnification and redshift.
Following the detailed analysis by \citet{Millon2022}, we assume that the microlensing signal is predominantly associated with image B, while Image A serves as the intrinsic quasar baseline. This assumption is supported by three key observational evidences: (1) image B shows a significant magnitude excess ($\sim 0.55$\,mag) compared to macro-model predictions; (2) image B shows a long-term fading trend inconsistent with that of image A; and (3) spectroscopic data reveal a low continuum-to-line contrast in image B, a signature characteristic of strong chromatic microlensing \citep{Faure09}.
We consider that the microlensing effect dominants the variation of the image B in the observation period, so that the image A after the correction of the macrolensing amplification mostly reflects the intrinsic spectrum of the quasar. 

\label{sec:J0158}

\begin{figure}
\centering
\includegraphics[width=0.46\textwidth]{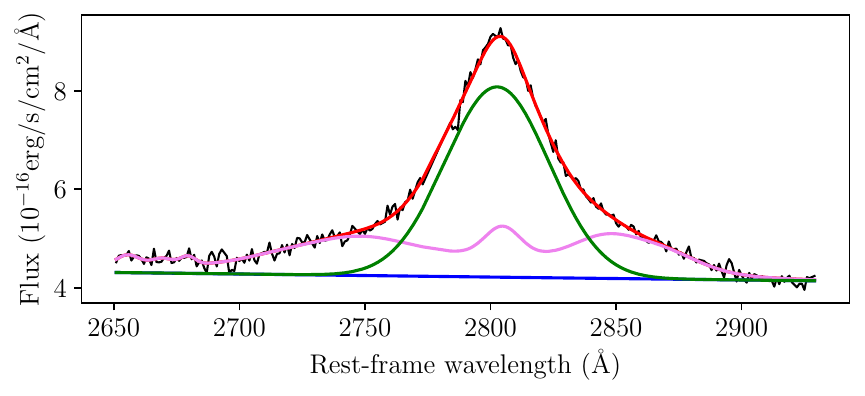}
\caption{
Decomposition of the profile of Mg II emission line. The observed spectrum (black line) of the Mg II profile is fitted by three components: a power-law continuum (blue line), Fe II lines (violet line), and Mg II broad emission line in Gaussian profile (green line), with the whole fitting model shown in red line. 
}
\label{fig:Mass}
\end{figure}

\begin{figure}
\centering
\includegraphics[width=0.46\textwidth]{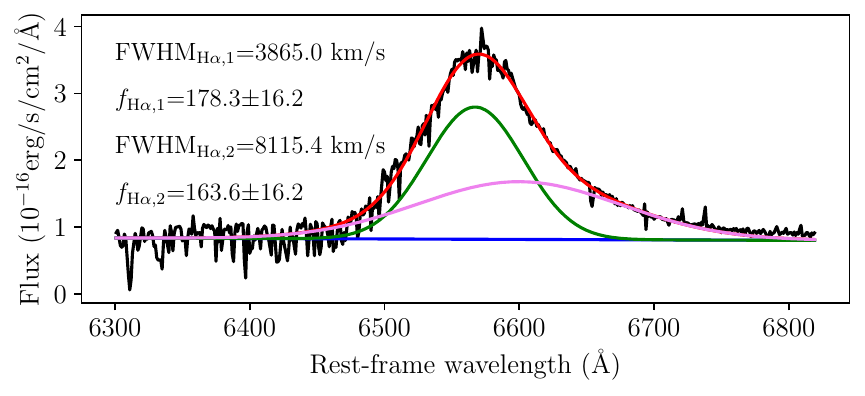}\\
\includegraphics[width=0.46\textwidth]{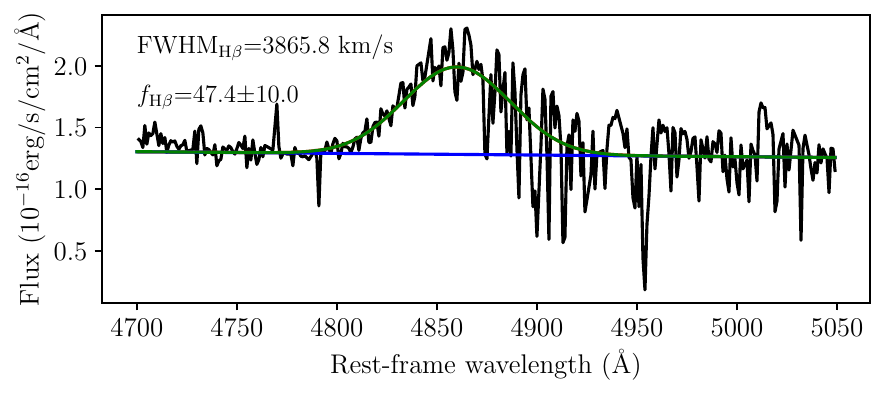}
\caption{
Profiles of the broad emission line H$\alpha$ (top panel) and H$\beta$ (bottom panel). The profile of BEL H$\alpha$ can be decomposed into two components, one with width $\sim3865~{\rm km~s^{-1}}$ and the other $8115~{\rm km~s^{-1}}$. However, H$\beta$ can only be fitted with a single component with width consistent with the less broad component of H$\alpha$, and the broader component, if any, cannot be extracted due to the low signal to noise ratio of H$\beta$ BEL at the red wing.
}
\label{fig:extinction}
\end{figure}

We extract broad emission lines Mg II, H$\alpha$, and H$\beta$ from the spectrum by fitting the continuum around each line as a power law. As shown in Figure~\ref{fig:Mass}, the profile of Mg II $\lambda 2798$ emission line is decomposed into three components: a power-law continuum (blue), Fe II pseudo-continuum from \citet{Sigut2003} templates (violet), and a broad Gaussian emission line (green). The measured full width at half maximum (FWHM) of Mg II ($5202\pm150 {\rm km s}^{-1}$) combined with the demagnified ($\mu_{{\rm macro},A}=2.23$) $3000\AA$ luminosity $L_{3000} = (5.74 \pm 0.12) \times 10^{45} {\rm erg~s^{-1}}$ yields an SMBH mass of $\log(M_\bullet/M_\odot) = 9.04 \pm 0.10 $ via the virial relation from \citet{Peng2006}:
\begin{equation}
M_\bullet = 6.1 \times 10^6 \left( \frac{\rm FWHM_{\rm Mg II}}{1000  {\rm km~s^{-1}}} \right)^2 \left( \frac{L_{3000}}{10^{44} {\rm erg~s^{-1}}} \right)^{0.47} M_\odot.
\end{equation}

Figure~\ref{fig:extinction} shows the decomposition of the broad emission line H$\alpha$ and H$\beta$.
While H$\alpha$ requires two broad components for optimal fitting, the lower signal-to-noise H$\beta$ profile only permits decomposition of the narrower component. The dust extinction inferred from the narrower component of H$\alpha$ and H$\beta$ BELs is $E(B-V)\sim 2.5/3.1\times \log(f_{\rm H\alpha}/f_{\rm H\beta}/2.86)=0.096$.

We obtained ultraviolet spectroscopic data for Q J0158-4315 from Hubble Space Telescope (HST) observations conducted under program ID 17146 (PI: Zahedy). The science-grade spectra were acquired using the Space Telescope Imaging Spectrograph (STIS) with the G140L grating, which covers the wavelength range of 1140-1730 Å. The observations specifically targeted J0158-4325AB and took place on August 16, 2023, with a total exposure time of 7298 seconds. The data were processed to calibration level 2 and retrieved from the Hubble Spectroscopic Legacy Archive (HASP). The spectrum is displayed in the bottom panel of Figure~\ref{fig:observation}. Additionally, it has been corrected for Galactic extinction, macrolensing magnification, and redshifted to the rest frame.

\section{Models}
\label{sec:model}

We present a unified theoretical framework for SMBBH systems with triple-disk accretion structures, synthesizing constraints from microlensing observations and multi-wavelength spectral analysis. 
Our model self-consistently connects three fundamental aspects: the geometric configuration of the SMBBH-disk system, multi-wavelength emission properties, and time-dependent microlensing signatures.

\subsection{Geometric configuration and accretion disk structure}
\label{sec:disk}

The system consists of a primary black hole of mass $ M_{\bullet,1} $ and a secondary companion 
$M_{\bullet,2} $ (mass ratio $ q = M_{\bullet,2}/M_{\bullet,1} $) separated by orbital semi-major axis $ a_{\rm BBH} $. During inspiral, the secondary opens a gap with characteristic width $ R_{\rm H} \approx a_{\rm BBH}(q/3)^{1/3}$ in the primary's accretion disk \citep{Paschalidis2021}. Outside this cavity, a circumbinary disk persists, while each SMBH maintains its own mini-disk fed by gas streams from the circumbinary disk \citep{Farris14,Dittmann2023}. The circumbinary disk's inner truncation radius can be estimated by $r_{\rm in,c}=\abbh/(1+q) + R_{\rm H}(q)$, where $R_{\rm H}$ is the Hill radius. The hydrodynamical simulations indicate that $ r_{\rm in,c} $ evolves from $ \sim 1.2a_{\rm BBH} $ for $ q=0.1 $ to $ \sim 2a_{\rm BBH} $ for equal-mass systems \citep{dAscoli2018,Siwek23}. We parameterize this as $ r_{\rm in} = f_{\rm in} a_{\rm BBH} $, where $ f_{\rm in} \in [1.1, 2] $. Mini-disks around each SMBH are truncated at $ f_R R_{\rm RL} $, where $ R_{\rm RL}(x) = 0.49a_{\rm BBH}x^{2/3}/[0.6x^{2/3} + \ln(1+x^{1/3})] $ defines the Roche lobe radius ($ x = q ~{\rm or}~ 1/q $ for secondary/primary components) and $ f_R \leq 1 $ is a scaling factor accounting for incomplete Roche lobe filling \citep{Eggleton83,Paschalidis2021}.

For an SMBBH-triple disk system, the continuum radiation may be estimated according to simple disk accretion models. In these models, key radiative parameters include inner disk radius $r_{\rm in}$ and Eddington ratio $f_{\rm E}$. For standard thin disks, observational data suggest that radiation efficiency $\epsilon \approx 0.1$ \citep[e.g.,][]{YT02, YL08, Shankar13} corresponds to moderately spinning black holes ($a\approx0.67$), yielding innermost stable circular orbit (ISCO) radii $ r_{\rm ISCO} \approx 3.5 r_{\rm g}$, where $r_{\rm g} = GM_{\bullet}/c^2$ is the gravitational radius. Therefore, we set $r_{{\rm in},i}=3.5r_{{\rm g},i}$ for two mini-disks in this paper, where $i$ denotes the primary and secondary mini-disks..

For simplicity, we consider the binary black hole (BBH) systems to be in circular orbits, with the SMBBH and their associated triple accretion disks lying in the same plane. We utilize the standard thin disk model \citep[e.g.,][]{SS73, NT73} to characterize the accretion flow around each supermassive black hole (SMBH) as well as the circumbinary disk. According to this model, the emission from the disk can be approximated as multicolor black body radiation, with the effective temperature at a given radius $r$ expressed as
\begin{equation}
\teff(r) = \left[\frac{3G\mbh \dot{M}_{\rm acc}}{8\pi\sigB r^3}\left(1-\sqrt{\frac{\rin}{r}}\right)\right]^{1/4},
\label{eq:temp}
\end{equation}
%
which can also be written as
\bea
\teff(r)&\simeq& 2\times 10^5~{\rm K} \left(\frac{0.1}{\epsilon}\right)^{1/4} \left(\frac{f_{\rm E}}{0.3}\right)^{1/4}  \nonumber \\
& & \times \left(\frac{10^8\msun}{M_{\bullet}}\right)^{1/4}
\left(\frac{\rin}{r}\right)^{3/4} \left(1-\sqrt{\frac{\rin}{r}}\right)^{1/4}.
\eea
In this equation, $G$ denotes the gravitational constant, $\sigma_B$ represents the Stefan-Boltzmann constant, $M_{\bullet}$ signifies the mass of the black hole, and $\dot{M}_{\rm acc}$ indicates the accretion rate. We define the Eddington ratio as $f_{\rm E}=\dot{M}_{\rm acc}/\dot{M}_{\bullet,\rm E}$. Here, the Eddington accretion limit is expressed as $\dot{M}_{\bullet,\rm E}=4\pi m_{\rm p}G M_{\bullet}/\epsilon c \sigma_{\rm T}$, with $c$ denoting the speed of light, $m_{\rm p}$ the proton mass, and $\sigma_{\rm T}$ the Thomson cross-section.

In the context of an SMBBH-triple disk system,we assume that the temperature profiles follow the standard thin disk model (as described in Eq.~\ref{eq:temp}). Thus, the monochromatic specific intensity of the thin disk at wavelength $\lambda$ and radius $r$ is given by
\begin{equation}
B_{\lambda}(r) = \frac{2 hc^2/\lambda^5}{\exp\left[\frac{\hp c}{\lambda k_{\rm B}\teff(r)}\right]-1},
\end{equation}
where $\hp$ is the Planck constant and $k_{\rm B}$ is the Boltzmann constant. The surface brightness profile for a narrowband filter centered at wavelength $\lambda$ is then described by
\begin{equation}
I(r) = \int_{\lambda_{1}}^{\lambda_{2}} B_{\lambda}(r) d\lambda.
\end{equation}
Each disk is truncated at its respective inner and outer boundaries. The outer radii for the primary and secondary mini disks are given by $ r_{\rm out,1} = f_R R_{\rm RL}(1/q) $ and $ r_{\rm out,2} = f_R R_{\rm RL}(q) $, respectively. Their inner radii are specified as $ r_{\rm in,1} = 3.5GM_{\bullet,1}/c^2 $ and $ r_{\rm in,2} = 3.5GM_{\bullet,2}/c^2 $. The circumbinary disk has an inner radius of $ r_{\rm in,c} = f_{\rm in}\abbh $ and an outer radius of $ r_{\rm out,c} = 1000r_{\rm g} =1000G(M_{\bullet,1}+M_{\bullet,2})/c^2 $. The Eddington ratios for the accretion onto each supermassive black hole (MBH) are represented as $ f_{\rm E,1} $ and $ f_{\rm E,2} $, while for the circumbinary disk, we define $ f_{\rm E,c} = \frac{f_{\rm E,1}}{1+q} + \frac{q f_{\rm E,2}}{1+q} $, assuming continuous accretion with $ \dot{M_c} = \dot{M_1} + \dot{M_2} $. Moreover, we utilize the empirical relationship $ \dot{M}_{\bullet,1} = \dot{M}_{\bullet,2}(0.1 + 0.9q) $ based on simulation findings reported in \citet{Duffell2020}.

Using this simplified single/triple-disk model, we can calculate the R-band brightness distribution for an active binary black hole (BBH) system with a triple disk configuration. To align with the observed light curve of Q J0158-4325 in the R band, we define the wavelengths as $\lambda_1 = \frac{6100}{1+z_{\rm obs}}\mathring{A}$ and $\lambda_2 = \frac{7200}{1+z_{\rm obs}}\mathring{A}$, where $z_{\rm obs}=1.29$ denotes the redshift.

\subsection{Multi-component SED}

The unique triple-disk structure of an active SMBBH system produces a SED that is distinct from that of a single active SMBH. A key characteristic, particularly for close binaries with separations on the order of (O($100r_{\rm g}$)), is a deficit of emission in the optical and UV bands.

The total flux density is the sum of contributions from the circumbinary disk and the two inner mini-disks. For a single, thin accretion disk, the observed flux density at wavelength $\lambda$ is given by:
\begin{equation}
F_\lambda=\frac{2 \pi \cos i}{D_L^2} \int_{r_{\rm in}}^{r_{\rm out}} B_\lambda(r) r \mathrm{~d} r,
\end{equation}
where $D_L$ is the luminosity distance to the source, and $i$ is the inclination angle of the disk relative to the observer's line of sight.
For simplicity, we assume a face-on orientation (
$i=0$, hence $\cos i=1$). By combining the emissions from the three disks, we derive the SED of the SMBBH system as follows:
\bea
\lambda L_{\lambda}&=&4\pi D^2 \lambda F_{\lambda}
%
=8\pi^2\lambda\left(\int^{r_{\rm out,c}}_{r_{\rm in,c}} B_{\lambda}(r)rdr+ \right. \nonumber\\
&&\left.\int^{r_{\rm out,1}}_{r_{\rm in,1}} B_{\lambda,1}(r)rdr
+\int^{r_{\rm out,2}}_{r_{\rm in,2}} B_{\lambda,2}(r)rdr\right).
\eea
Here, the subscripts $c$, 1 and 2 denote the circumbinary disk, the mini-disk of the primary black hole, and the mini-disk of the secondary black hole, respectively. In this formulation, we are only considering the emission from thin disks in the optical and UV bands.

\subsection{Microlensing light curves}
\label{sec:mocklc}

The effect of microlensing on a macroimage is illustrated through a magnification map in the source plane. Each pixel in this map represents the magnification of the macroimage relative to its average value when a point source is located at that specific point. These maps are generated using the ray-shooting technique, which models the deflection of light by a distribution of stellar lenses within the galaxy, characterized by parameters such as the surface mass density ($\kappa$) and shear ($\gamma$) \citep{Kochanek04, Wambsganss06, Kayser86, Paczynski86, Wambsganss90, WPS90}.

In our simulations, we adopt the fixed lens parameters ($\kappa$, $\gamma$) = (0.72, 1.03), with the stellar contribution set to $\kappa_*/\kappa=0.92$. These values are consistent with the macrolensing model for image B of Q J0158-4325 \citep{Millon2022}. The stars are randomly positioned, and their masses are drawn from a Salpeter initial mass function, $dN/dm\propto m^{-2.35}$, within the mass range of 0.1 to 10 $\msun$. For these parameters, the Einstein radius of a star in the source plane is $\left< \re \right> = 3.4 \times 10^{16} \left(\left< m_* \right> / 0.3 \msun \right)^{1/2}$ cm.

Since quasars are extended sources, we convolve the original magnification maps with a model of the source's surface brightness distribution. We model the source as a SMBBH system. Its time-dependent surface brightness distribution is governed by the orbital motion of the binary and its associated disks around the center of mass, with an orbital period $P_{\rm orb}$. The rotation of this triple-disk system creates a dynamically changing brightness profile, particularly within the central cavity.
 
To generate a microlensing light curve, we define a source trajectory across the magnification map, which represents the effective relative motion between the source and the lens plane with a velocity $v_{\rm e}$. At each observational epoch corresponding to a point on this trajectory, we convolve the instantaneous surface brightness map of the SMBBH with the underlying magnification map. The measured magnitude at epoch $i$ is then given by:
\begin{equation}
m_i = -2.5 \log \mu_{{\rm micro},i} - 2.5 \log |\mu_{{\rm macro},i}| + m_{\rm int},
\end{equation}
where $\mu_{{\rm micro},i}$ is the microlensing amplification (encoding small-scale fluctuations from stellar lenses), $\mu_{{\rm macro},i}$ is the smooth macrolensing amplification from the galaxy-scale potential, and $m_{\rm int}$ is the intrinsic magnitude of the source.

For systems with multiple images, intrinsic source variations can be isolated and removed by leveraging the known time delays and magnifications between images. As this work focuses purely on microlensing, we isolate the microlensing-induced deviations by subtracting the mean magnitude. Assuming the macrolensing magnification and intrinsic magnitude are constant over the observing period, the magnitude deviation simplifies to
\begin{equation}
\delta m_i = m_i - \left\langle m_i \right\rangle = -2.5 \log\left(\mu_{{\rm micro},i}\right).
\end{equation}
The timescale of these $\delta m_i$ variations is set by the effective velocity $v_{\rm e}$, which we adopt as a fixed parameter for each simulated light curve.

\section{Results}

Building on previous findings that SMBBH systems exhibit distinct periodic behaviors depending on their mass ratios, we aim to test whether binary orbital motion can account for the observed microlensing variability. In unequal-mass systems ($q\ll 1$), periodic variations arise from the displacement of the brighter accretion disk (primary) relative to the center of mass. Conversely, in near-equal-mass systems ($q \sim 1$), both mini-disks contribute comparably to the emission, producing variability with a period half the orbital period of the secondary black hole. To evaluate these scenarios, we compare two orbital period models: one where the orbital period matches the observed variability period ($P_{\rm orb} = P_{\rm obs}$) and another where it is twice the observed period ($P_{\rm orb} = 2P_{\rm obs}$). The orbital period of a binary system in rest-frame is determined by the total mass $M_{\bullet}$ and the separation $a_{\rm BBH}$ according to the formula: $P_{\rm orb}=2\pi (a^3_{\rm BBH}/GM_{\bullet})^{1/2}$. Thus given the orbital period, we can compute the corresponding separation:
\bea
a_{\rm BBH} = \left(\frac{GM_{\bullet} P_{\rm orb}^2}{4\pi^2}\right)^{1/3}
=43 r_{\rm g} \left(\frac{M_{\bullet}}{10^9\msun}\right)^{-2/3}\left(\frac{P_{\rm orb}}{100{\rm day}}\right)^{2/3},
\label{eq:abbh}
\eea
where $r_{\rm g} = GM_{\bullet}/c^2$ is the gravitational radius. In the following fitting of the microlensing light curve and SED, we considered this two orbit period models and provided the $a_{\rm BBH}$ based on the orbit period and the total black hole mass.

\begin{table*}
\centering
\caption {Parameters of the different cases. Here $P_{\rm orb}$ is the orbit period in observer frame, $f_R$ is the fraction of actual disk size of the mini disk to the Roche lobe radius. $r_{\rm out,1}$ and $r_{\rm out,2}$ are the physical disk size of the primary and secondary mini disks. And $\tau_{\rm c}$ is the coalescence time in observer frame of the SMBBH due to the gravitational wave decay. Here we set circumbinary disk's inner trunction radius $r_{\rm in}=\abbh/(1+q) + R_{\rm H}(q)$.}
\begin{tabular}{cccccccccccc} \hline\hline
Model & $M_{\bullet}(\msun)$ & $q$ & $f_{\rm E,1}$ & $f_{\rm E,2}$ & $f_{\rm E,c}$ &$P_{\rm orb}$(days)& $\abbh(r_{\rm g})$&$f_R$ &$r_{\rm out,1}$(cm)&$r_{\rm out,2}$(cm)&$\tau_{\rm c}$(yr)\\ \hline
S0&$1.11\times 10^9$&&&&0.1&&&&&&\\
M1 & $1.11\times10^9$ & 0.1&0.018   & 0.92  & 0.1 &172.6 &33  & 0.5  &$10^{15.2}$&$10^{14.7}$&111\\
M2 & $1.11\times10^9$ & 0.1 &0.018  & 0.92  & 0.1   &172.6 &33  &1.0&$10^{15.5}$&$10^{15.0}$&111\\
M3 & $1.11\times10^9$ & 0.5&0.053   & 0.19  & 0.1 &172.6 &33  & 0.5 &$10^{15.1}$&$10^{14.9}$&41\\
M4 & $1.11\times10^9$ & 0.5 &0.053  & 0.19  & 0.1  &172.6  &33  &1.0&$10^{15.4}$&$10^{15.2}$&41\\ \hline
M5 & $1.11\times10^9$ & 0.5&0.053& 0.19  & 0.1 &345.3 &52  & 0.5  &$10^{15.3}$&$10^{15.1}$&263\\
M6 & $1.11\times10^9$ & 0.5 &0.053  & 0.19  & 0.1 &345.3   &52  &1.0&$10^{15.6}$&$10^{15.4}$&263\\
M7 & $1.11\times10^9$ & 1.0& 0.1  & 0.1  & 0.1   &345.3  &52 &0.5 &$10^{15.2}$&$10^{15.2}$&233\\
M8 & $1.11\times10^9$ & 1.0& 0.1  & 0.1  & 0.1   &345.3  &52 &1.0 &$10^{15.5}$&$10^{15.5}$&233\\ \hline\hline
\end{tabular}
\label{tab:t1}
\end{table*}

\subsection{Fitting the microlensing lightcurve}

\begin{figure}
\centering
\includegraphics[width=0.3\textwidth,angle=-90]{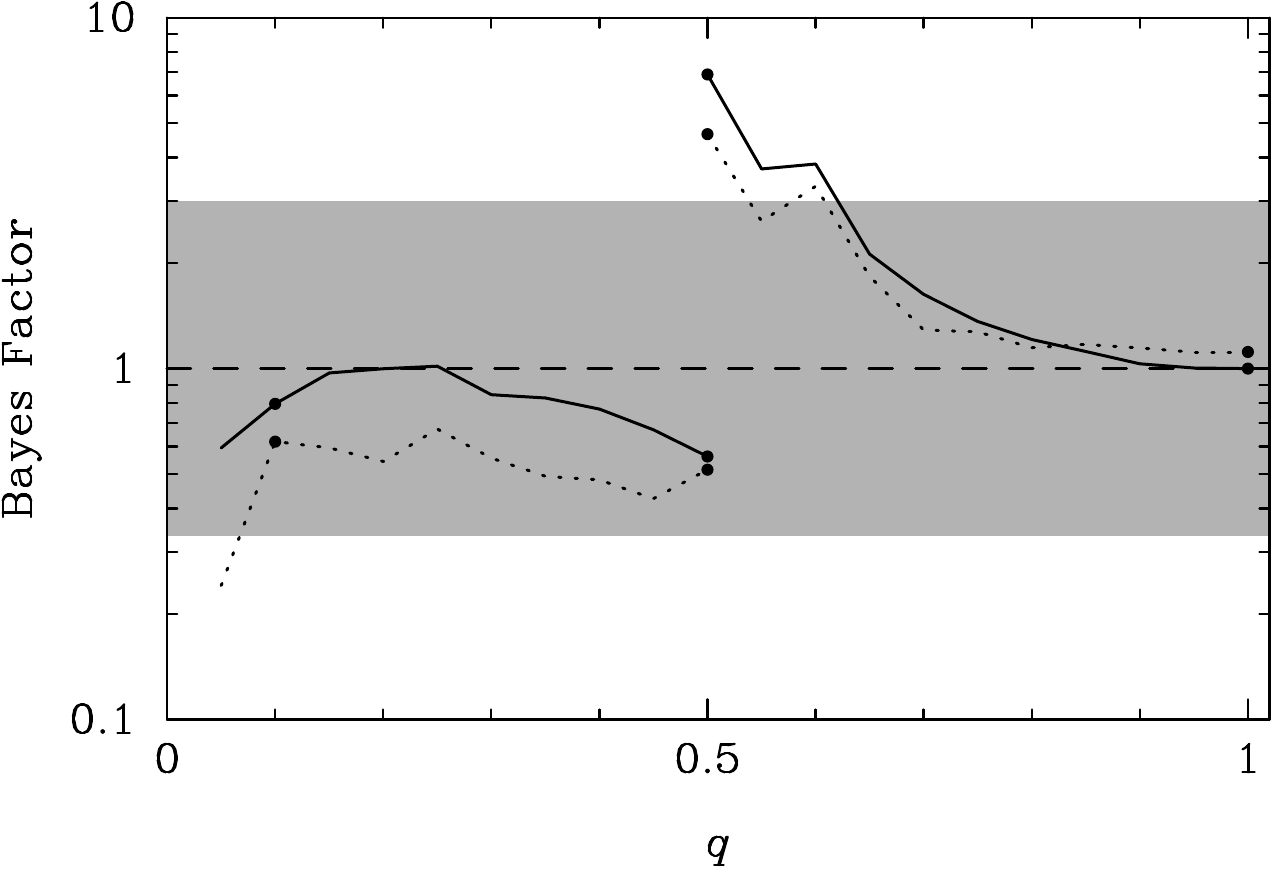}
\caption{
Bayes factors for light curve fitting, with black hole masses and circum-binary disk accretion rates fixed at $f_{\rm E,c}=0.1$. Solid lines correspond to $f_R=1.0$, while dotted lines represent $f_R=0.5$. 
For mass ratios $q>0.5$ we set the orbital period $P_{\rm orb}=2P_{\rm obs}$; for $q<0.5$, we set $P_{\rm orb}=P_{\rm obs}$. The dashed horizontal line marks a Bayes factor of unity, with the shaded gray area representing Bayes factors between $\frac{1}{3}$ and $3$. The points correspond to models M1-M8, as detailed in Table \ref{tab:t1}. Here we fixed the inner radius of the circumbinary disk as $r_{\rm in}=\abbh/(1+q) + R_{\rm H}(q)$.
}
\label{fig:BFl}
\end{figure}

\begin{figure*}
\centering
\includegraphics[width=0.76\textwidth,angle=-90]{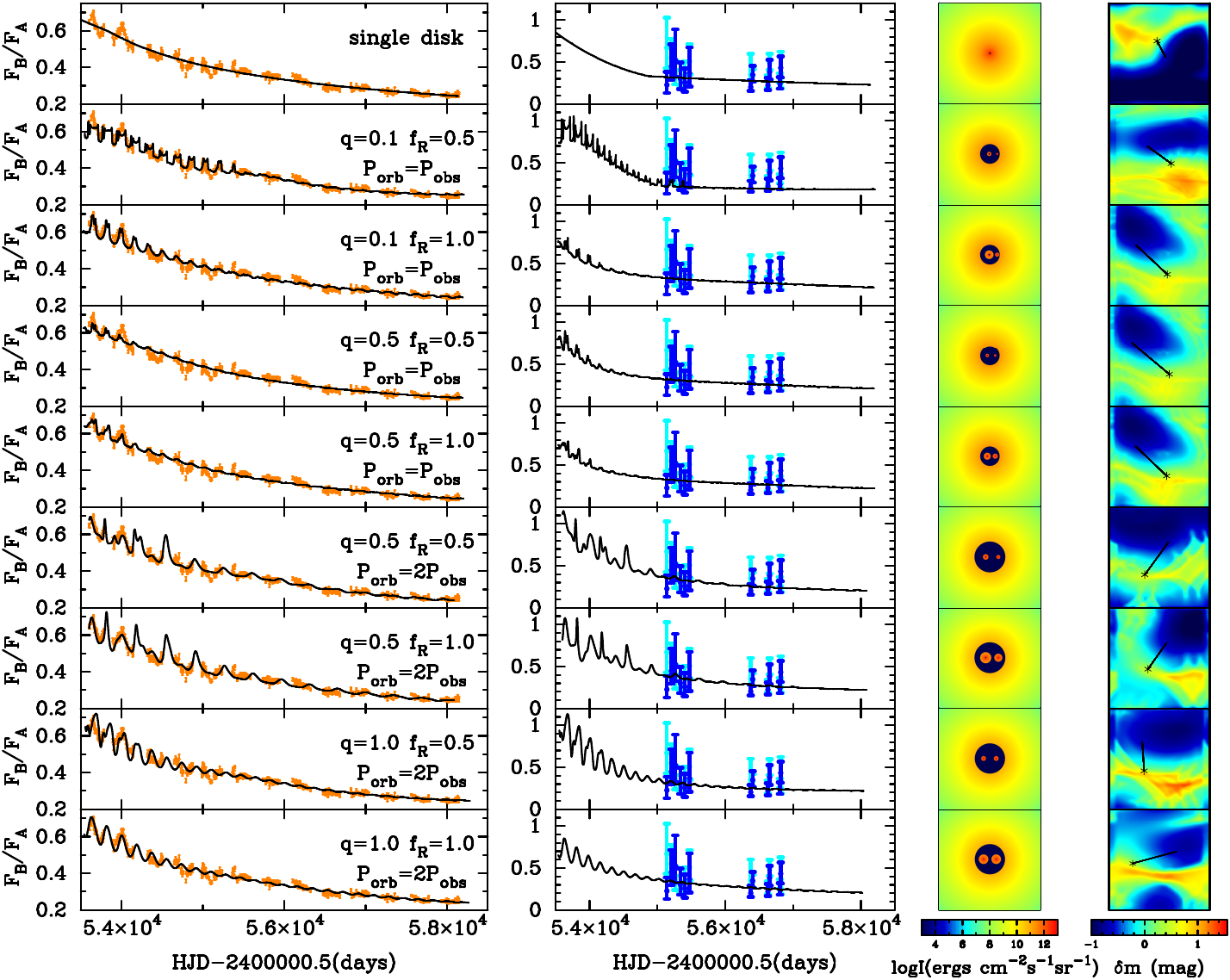}
\caption{
Left panel: Flux ratio $F_B/F_A$ observed by the Euler telescope from 2005 to 2018 (orange dots with error bars), with the solid lines representing the best-fit model for different cases. Middle left panel: Blue (cyan) dots with errors are the soft (hard) X-ray flux ratio $F_B/F_A$ as observed by Chandra over the period 2009-2014 and the black solid lines are corresponding microlensing light curve prediction in the X-ray band. Middle right panel: Surface brightness distribution map of SMBBH system at a given time. Each map has a side length of 2 times the Einstein radius ($R_{\rm E}$) and consists of $2048\times2048$ pixels. Right panel: Convolution microlensing magnification map, with the black line indicating the trajectory that best fits the microlensing light curve. The maps from top to bottom correspond to the S0 and M1-M8 case. 
}
\label{fig:allmass}
\end{figure*}

To simplify our analysis of the microlensing light curve, we will adopt the following fixed parameters: the total black mass $\mbh=1.1\times10^9\msun$, derived from Mg II line measurements; the Eddington ratio of circumbinary disk $f_{E,c}=0.1$; and accretion rates for the mini-disks, following the relation $\dot{M}_{\bullet,1} = \dot{M}_{\bullet,2}(0.1 + 0.9q)$. This empirical formula accounts for the asymmetric mass supply to the primary and secondary black holes in simulations.

As described in Section~\ref{sec:disk} we create triple disk surface brightness profile. We begin by generating magnification maps for the image B in the source plane, following the methodology outlined in Section~\ref{sec:mocklc}. These maps are then convolved with the source's surface brightness distribution. We simulate a range of light paths by randomly selecting both starting points and directions on the convolved magnification map. At each given time point, we calculate the microlensing-induced magnitude variation, $\delta m_{i,B}$, by convolving the time-dependent magnification map with the source's surface brightness distribution. This process accounts for changes in the magnification map due to the rotation of the SMBBH-triple disk system. From these calculations, we construct a series of model light curves, $\delta m'_i$, for each trajectory.

Assuming that the image A is not influenced by microlensing, i.e., $\mu_{{\rm micro},A}=1$, and the time delay between the images A and B is subtracted, 
then the magnitude difference between the two images is caused by the microlensing effect on the image B and the difference in the macrolensing magnification of the two images, i.e., 
\bea
\Delta m_{\rm AB} &=& m_{\rm B} - m_{\rm A}\nonumber\\
&=& -2.5 \log \mu_{\rm micro,B} - 2.5 \log \left|\frac{\mu_{\rm macro,B}}{\mu_{\rm macro,A}}\right|.
\eea
To fit the model light curves to the observed data, we employ the standard $\chi^2$ statistic:
\be
\chi_{\rm l}^2= \sum_i \frac{ \left( \delta m_i+\Delta m_0 - \delta m_{{\rm o},i} \right)^2}{\sigma_{i}^2},
\ee
where $\delta m_{\rm o,i}=m_{\rm o,B}-m_{\rm o,A}$ represents the observed magnitude difference between the images A and B, and $\sigma_{i}$ denotes the observational errors. The term $\delta m_i=-2.5 \log \mu_{\rm micro,B}$ corresponds to the modeled microlensing magnitudes. To obtain the best fit, we adjust the effective relative velocity $v_{\rm e}$, the initial orbital phase, and the trajectories. By minimizing the $\chi^2$ value, we identify the best-fitting light curve.

Initially, we investigate a wide area of the magnification map, covering $20\left<\re \right>\times 20\left<\re \right>$, to perform low-resolution light curve fitting. We randomly select $10^7$ trajectories and assign a value of $v_e$ to each model, subsequently calculating the $\chi_{\rm l}^2$ for each trajectory. To compare the models with different parameters, we define a Bayes factor as $\mathrm{BF}_l=\frac{\mathcal{Z}_8}{\mathcal{Z}}$ where $\mathcal{Z}$ is the evidence for the model with a mass ratio of $q$ and $f_{\rm R}$ and can be calculated by
\begin{equation}
\mathcal{Z}=\int \mathcal{L}\left(D \mid \theta, {\rm M}_i\right) p\left(\theta \mid {\rm M}_i\right) d \theta.
\end{equation}
Here $\theta$ is the parameters exclued $q$ and $f_{\rm R}$, the likelihood can be estimated as
$\mathcal{L}\left(D \mid \theta, {\rm M}_i\right)\propto e^{-\frac{1}{2}\chi^2}$
and $p(\theta\mid {\rm M}_i)$ is the prior distribution.
We add all the likelihood of $10^7$ different trajectories to represent the integral. And $\mathcal{Z}_8$ here represent the evidence of model M8 which list in Table~\ref{tab:t1}. A Bayes factor $\mathrm{BF}_l > 3$ indicates strong evidence for model M8 over the compared model, while $\mathrm{BF}_l < 1/3$ favors the alternative model. Values between $1/3$ and $3$ suggest comparable performance. Figure~\ref{fig:BFl} shows the low resolution fitting Bayes factor with different parameters. For $P_{\rm orb} = P_{\rm obs}$ case all the Bayes Factor are between $\frac{1}{3}$ to 3, this means all these models are comparable with model M8. For $P_{\rm orb} =2P_{\rm obs}$ case when the mass ratio is smaller then 0.5 the evidence is much smaller then M8 because its Bayes factor is larger then 3 and we think these models cannot fitting the microlensing light curve well. Then we choose eight typical model and refine our search to a smaller $2\left<\re \right>\times 2\left<\re \right>$ area to obtain a high-resolution fitting. Here we list the parameters of them in Table~\ref{tab:t1}. Figure~\ref{fig:allmass} presents the high resolution light curve fitting results for Q J0158-4325. As shown in these figures, for the single-black-hole disk model S0, it fails to reproduce the observed periodic fluctuation of the light curve and the Bayes factor comparing our fiducial SMBBH model (M8) to the single-black-hole disk model (S0) is $3.4$, indicating positive evidence in support of the SMBBH scenario.
For the $P_{\rm orb} = P_{\rm obs}$ case (M1-M4) model M2 give a relatively better fit (especially for first several periods), although M1-M4 have similar evidence. In the case of $P_{\rm orb} =2P_{\rm obs}$ (M5-M8), due to the missing data for certain time periods, the $q=0.5$ model can still provide a relatively decent fit. However, we observe that the light curve displays a higher peak followed by a lower peak, corresponding to the secondary minidisk and the primary disk, respectively. In summary, we cannot provide a strong constraint on the mass ratio of Q J0158-4325 solely based on the observation light curve data.

\subsection{Fitting the SED}
\label{sec:result}

\begin{figure}
\centering
\includegraphics[width=0.32\textwidth,angle=-90]{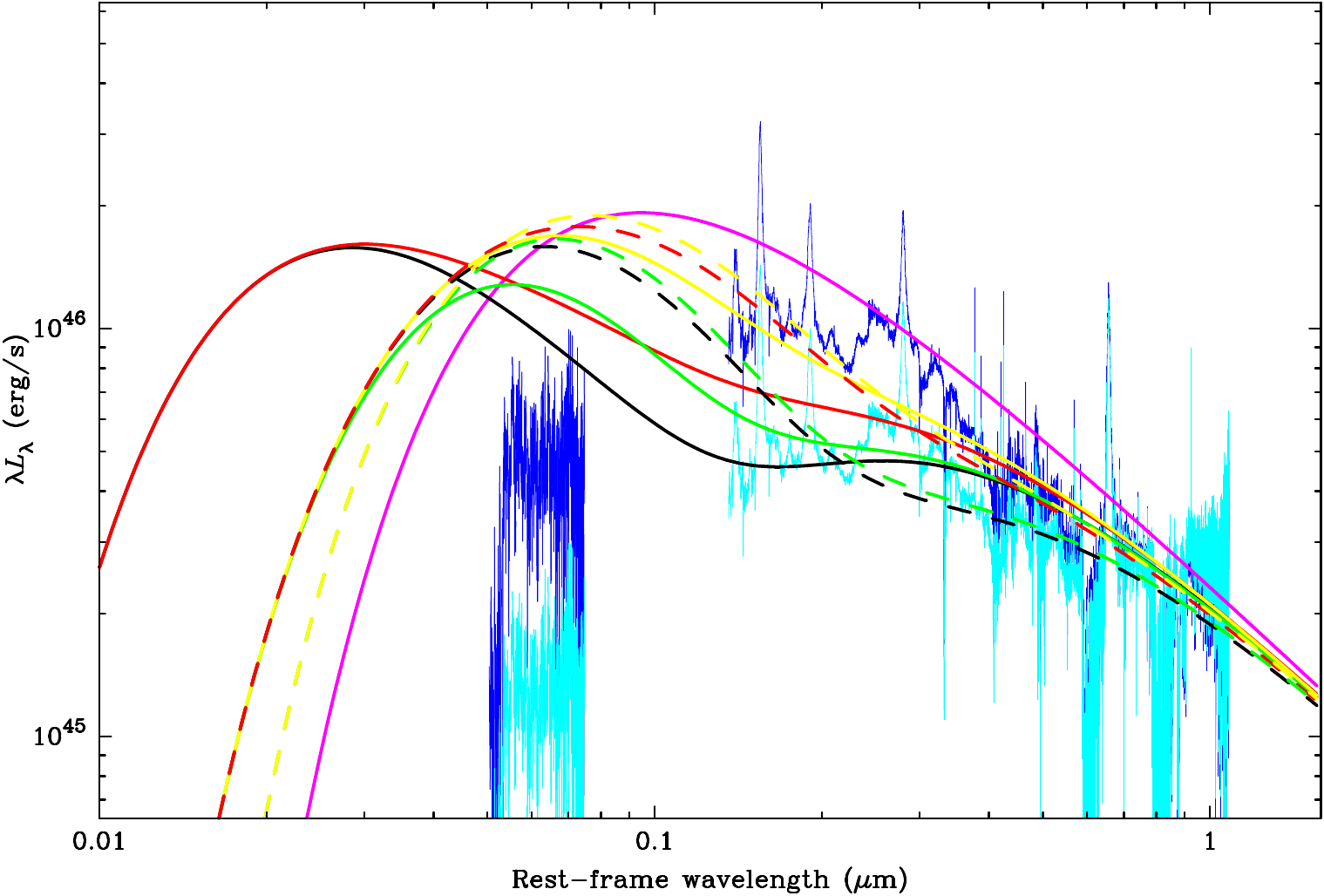}
\caption{
The solid lines represent the theoretical SEDs for models S0, M1, M3, M5, and M7, while the dashed lines depict the theoretical SEDs for models M2, M4, M6, and M8 listed in Table.\ref{tab:t1}. The pink, black, red, green, and yellow lines correspond to the model S0, M1(M2), M3(M4), M5(M6), and M7(M8), respectively.
The cyan line represents the macrolensing and redshift-corrected spectrum of Q J0158-4325, as observed with XSHOOTER and HST. In contrast, the blue line depicts the reddening-corrected spectrum, which has been adjusted for an extinction value of $E(B-V) = 0.096$.
}
\label{fig:SED}
\end{figure}
\begin{figure}
\centering
\includegraphics[width=0.55\textwidth,angle=-90]{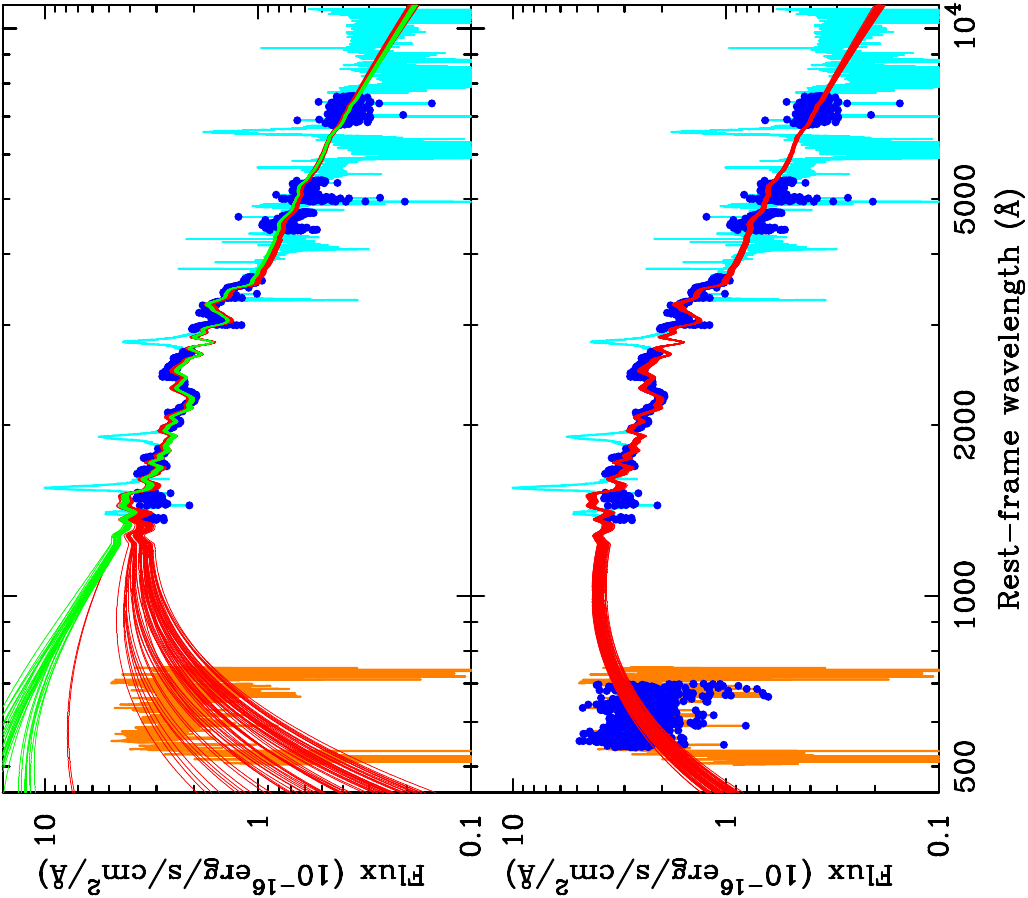}
\caption{
Comparison of the Spectrum of Q J0158-4325 with the the optimal SMBBH-triple disk model. The cyan and orange lines represent the macrolensing redshift-corrected spectra of Q J0158-4325, as observed by XSHOOTER and the HST, respectively. The blue dots indicate the data points utilized for model fitting. Top Panel: This panel includes only optical data for fitting, with all models shown having a reduced chi-squared value $\chi_{\rm s}^2/N_{\rm dof} < 1.45$. The red lines represent models with an orbital period $P_{\rm{orb}} = 2P_{\rm{obs}}$, while the green lines illustrate models with an orbital period $P_{\rm{orb}} = P_{\rm{obs}}$. Bottom Panel:  Both optical and UV data are employed for fitting. The lines depict the model that achieves a reduced chi-squared value of $\chi_{\rm s}^2/N_{\rm dof} < 1.65$. Notably, only models with $P_{\rm{orb}} = 2P_{\rm{obs}}$ demonstrate such a low chi-squared value.
}
\label{fig:SEDall}
\end{figure}

Figure~\ref{fig:SED} shows the SEDs from SMBBH-triple disk systems with different settings of the model parameters (see Table \ref{tab:t1}.) and the observation optical spectra of Q J0158-4325 with a correction of the macrolensing magnification and reddening. We can see that, although different models may provide similar microlensing light curves, such as M2 and M8 in Figure \ref{fig:allmass}, their SEDs are very different, as shown in Figure \ref{fig:SED}. Therefore, fitting the spectrum can provide further constraints on the model parameters.

The observed spectrum may be affected by the Fe\,II pseudo-continuum in the UV-optical range, which complicates the direct comparison between the observed spectrum and the one derived from the SMBBH-triple disk model. To address this, we include the Fe emission when fitting the SED of J0158-4325. We employ the template-fitting methodology proposed by \cite{Phillips77}, which is a commonly used approach, to account for the Fe emission. Specifically, we use the Fe spectrum from the narrow-line Seyfert 1 galaxy I Zw 1 to create an Fe II template. For the UV range, we combine the UV Fe templates established by \cite{Vestergaard01} and \cite{Tsuzuki06}. In the optical range, we construct the Fe template based on the list of Fe lines for I Zw 1 provided by \citet{veron04}. We then convolve the template with a Gaussian function of a specified FWHM and scale to match the observed data. In the above algorithm, we assume that the Fe emissions in the UV and optical bands have the same width and there is no shift between the redshifts of the UV and the optical lines. We also introduce a parameter that describe the ratio of the UV Fe flux to the optical Fe flux ($I_{\rm Fe,UV}/I_{\rm Fe,opt}$). For simplicity, we fixed the FWHM of the iron lines to $5202\rm{km s}^{-1}$, which corresponds to the line width of Mg II. 

We employ the Markov Chain Monte Carlo (MCMC) technique to sample the posterior distributions of the model parameters, specifically using the public Python package emcee \citep{Foreman-Mackey2013}, which implements the Affine Invariant Ensemble Sampler proposed by \citet{Goodman2010}. We initialize $100$ walkers for parameter space exploration, run the chain for $20,000$ steps, and discard the initial $15,000$ steps as burn-in to ensure the chain had converged to the stationary distribution. Posterior constraints and corner plots are subsequently generated from the remaining samples.

We use eight parameters to fit the spectrum, which are total black hole mass($M_{\bullet}$), mass ratio ($q$), Eddington ratio of secondary black hole ($f_{\rm E,2}$), dust extinction ($E(B-V)$), the scale factor of iron lines, the ratio of UV Fe flux to the optical Fe flux, the parameters $f_R$ and $f_{\rm in}$ which describe the size of the minidisk and the inner radiu of circumbinary disk. As described in Section \ref{sec:disk}, we assume a continuous accretion process and apply an empirical relationship to determine the Eddington ratios of the primary disk ($f_{E,1}$) and the circumbinary disk ($f_{\rm E,c}$) based on the mass ratio $q$ and the Eddington ratio of the secondary disk ($f_{\rm E,2}$).

\begin{figure}
\centering
\includegraphics[width=0.5\textwidth,angle=0]{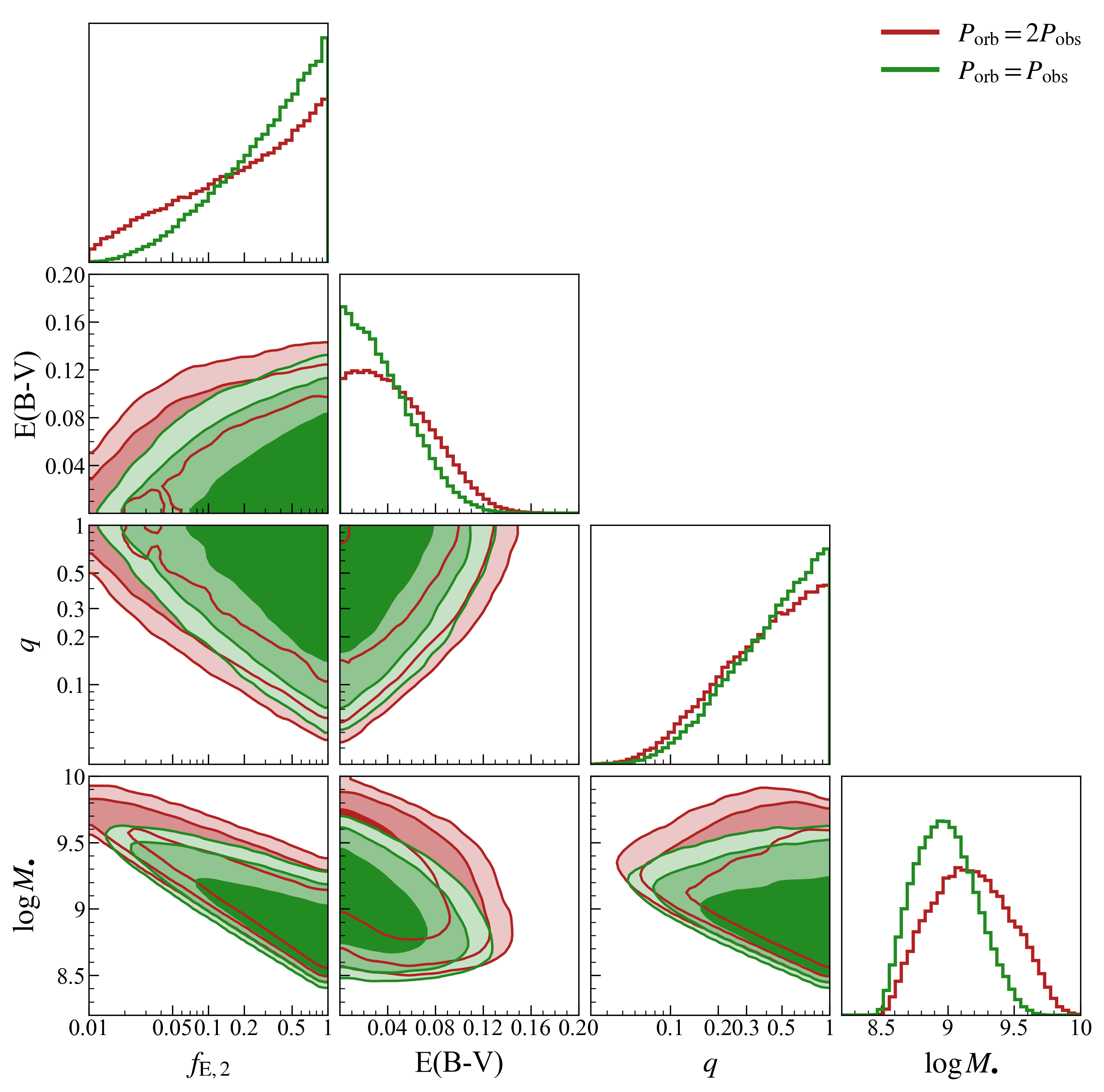}
\includegraphics[width=0.5\textwidth,angle=0]{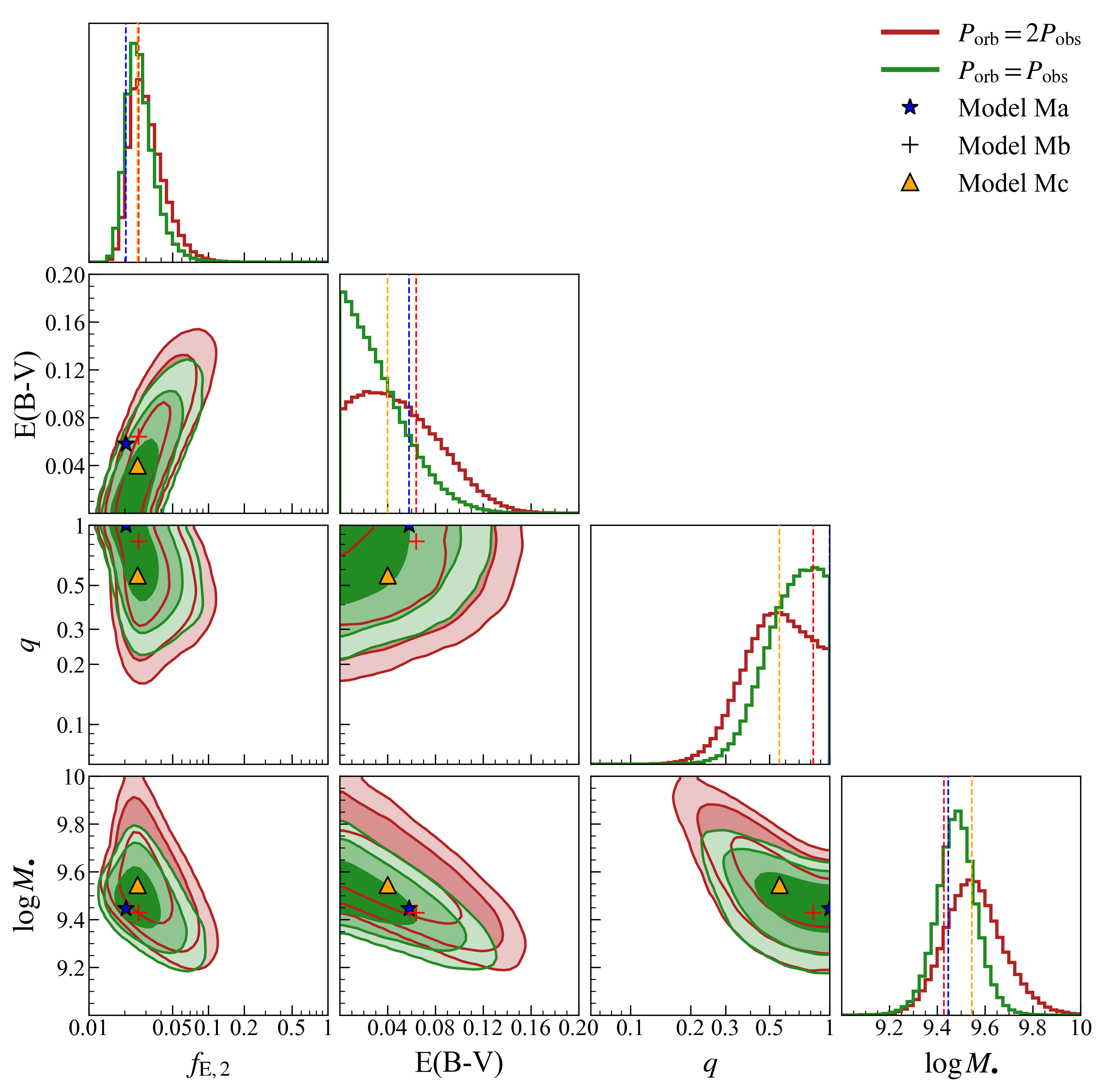}
\caption{
Posterior probability distributions for the parameters of the SMBBH system. The top panel presents results from fitting the optical spectrum alone, and the bottom panel shows results from a simultaneous fit of the optical and UV spectra. Red lines denote models with $P_{\rm orb} = 2P_{\rm obs}$, while green lines represent models with $P_{\rm orb} = P_{\rm obs}$. The blue star, red cross and yellow triangle correspond to Models Ma, Mb, and Mc in Table~\ref{tab:t2}, respectively.
}
\label{fig:disall}
\end{figure}

We chose a number of fitting windows in order to avoid strong emission or absorption lines, and the data used for the fitting are represented as blue dots in Figures~\ref{fig:SEDall}. First we only fitting the optical data from XSHOOTER which are shown in the top panel of Figure~\ref{fig:SEDall} with blue dots. Based on the information regarding the period obtained from the microlensing light curve, the system may either have a low mass ratio with an orbital period of $P_{\rm orb} = P_{\rm obs}$ or a high mass ratio with $P_{\rm orb}=2P_{\rm obs}$, so we consider this two different models and used a uniform prior for $\log q$ on the interval [-1.5,0]. With a fixed $P_{\rm orb}$ and the total black hole mass we can obtain $\abbh$. 
We present the model with a reduced chi-squared value of $\chi_{\rm s}^2/N_{\rm dof} < 1.5$ 
in the top panel of Figure~\ref{fig:SEDall}. Both the $P_{\rm orb} = P_{\rm obs}$ model and $P_{\rm orb}=2P_{\rm obs}$ model fit the optical spectrum effectively; however, their behaviors in the UV band differ significantly. The radiation in the UV band which the $P_{\rm orb} = P_{\rm obs}$ models predicted is much higher than the observation data. 

\begin{figure}
\centering
\includegraphics[width=0.3\textwidth,angle=-90]{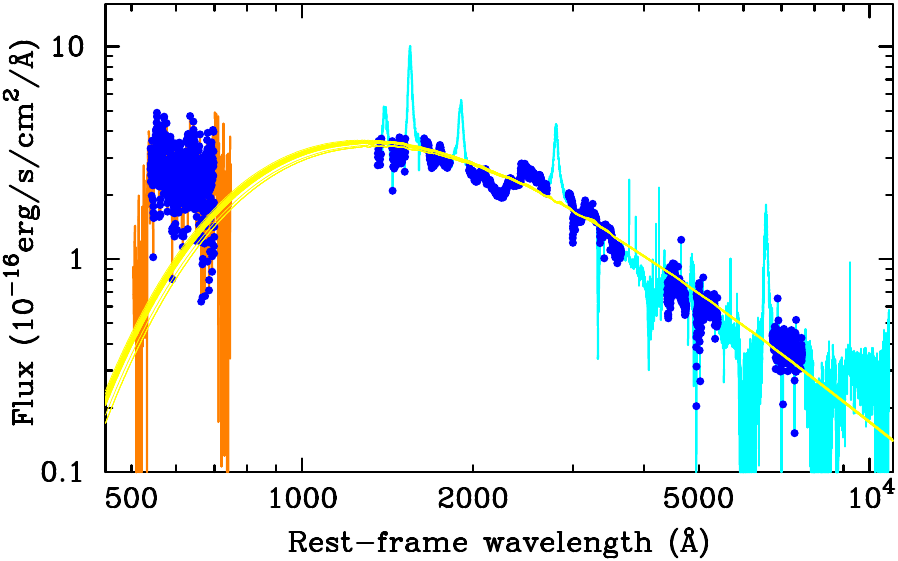}
\caption{
Comparison of the  Q J0158-4325 spectrum with the single-black-hole disk model. Optical and UV data are both used for the fitting. Yellow lines depict the model that achieves a reduced chi-squared value of $\chi_{\rm s}^2/N_{\rm dof} < 4.0$.
}
\label{fig:SEDsd}
\end{figure}

So then we fit both the optical and UV spectra simultaneously. The posterior distributions of the parameter for optical-only fitting and UV plus optical fitting are illustrated in Figure~\ref{fig:disall}. Both models exhibit a preference for a high mass ratio; however, the black hole mass in the $P_{\rm orb}=P_{\rm obs}$ model is lower. This is primarily due to the spectral fitting process, which tends to yield similar values for $a_{\rm BBH}$. Consequently, a smaller orbital period results in a reduced total black hole mass. We cannot directly compare the two models based on their posterior distributions. To assess the relative evidence for each model, we introduce a Bayesian factor as $\mathrm{BF}_s=p(D|M_{P_{\rm orb}=2P_{\rm obs}})/p(D|M_{P_{\rm orb}=P_{\rm obs}})$, where $p(D|M)$ is the marginal likelihood. Through MCMC simulation of the optical and UV spectra, we compute a Bayes factor of $\mathrm{BF}_s=4.0$. 
This provides substantial evidence for the $P_{\rm orb}=P_{\rm obs}$ model over the $P_{\rm orb}=2P_{\rm obs}$ model. However, when we incorporate physically motivated constraints—requiring $q<0.5$ in the $P_{\rm orb}=P_{\rm obs}$ model and $q>0.5$ for the $P_{\rm orb}=2P_{\rm obs}$ model the Bayes factor increases to $\mathrm{BF}_s=14$. 
This indicates strong evidence for the $P_{\rm orb}=2P_{\rm obs}$ model with a high mass ratio ($q>0.5$). In bottom panel of Figure~\ref{fig:SEDall}, we present the models which has a reduced chi-squared value of $\chi_{\rm s}^2/N_{\rm dof}< 1.65$ when fitting the optaical and UV data simultaneously. Our analysis indicates that all these are $P_{\rm orb}=2P_{\rm obs}$ model and have a mass ratio $q > 0.5$, allowing us to confidently rule out any models with a mass ratio less than 0.5. 
We also show the single-black-hole disk model fitting result in Figure~\ref{fig:SEDsd}. The best fit  yields a reduced $\chi^2$ of $\sim 4.0$, which is significantly poorer than the best fit to the observed spectrum by the SMBBH model. 

From Figure~\ref{fig:BFl}, we observe that for the case where $P_{\rm orb}=2P_{\rm obs}$, the evidence from the light curve fitting suggests a preference for a larger mass ratio. Therefore, by combining the evidence from both the microlensing light curve and the spectrum, we can infer that this SMBBH system is likely to have nearly equal black hole masses.

\subsection{Most likely models}

\begin{figure}
\centering
\includegraphics[width=0.35\textwidth,angle=-90]{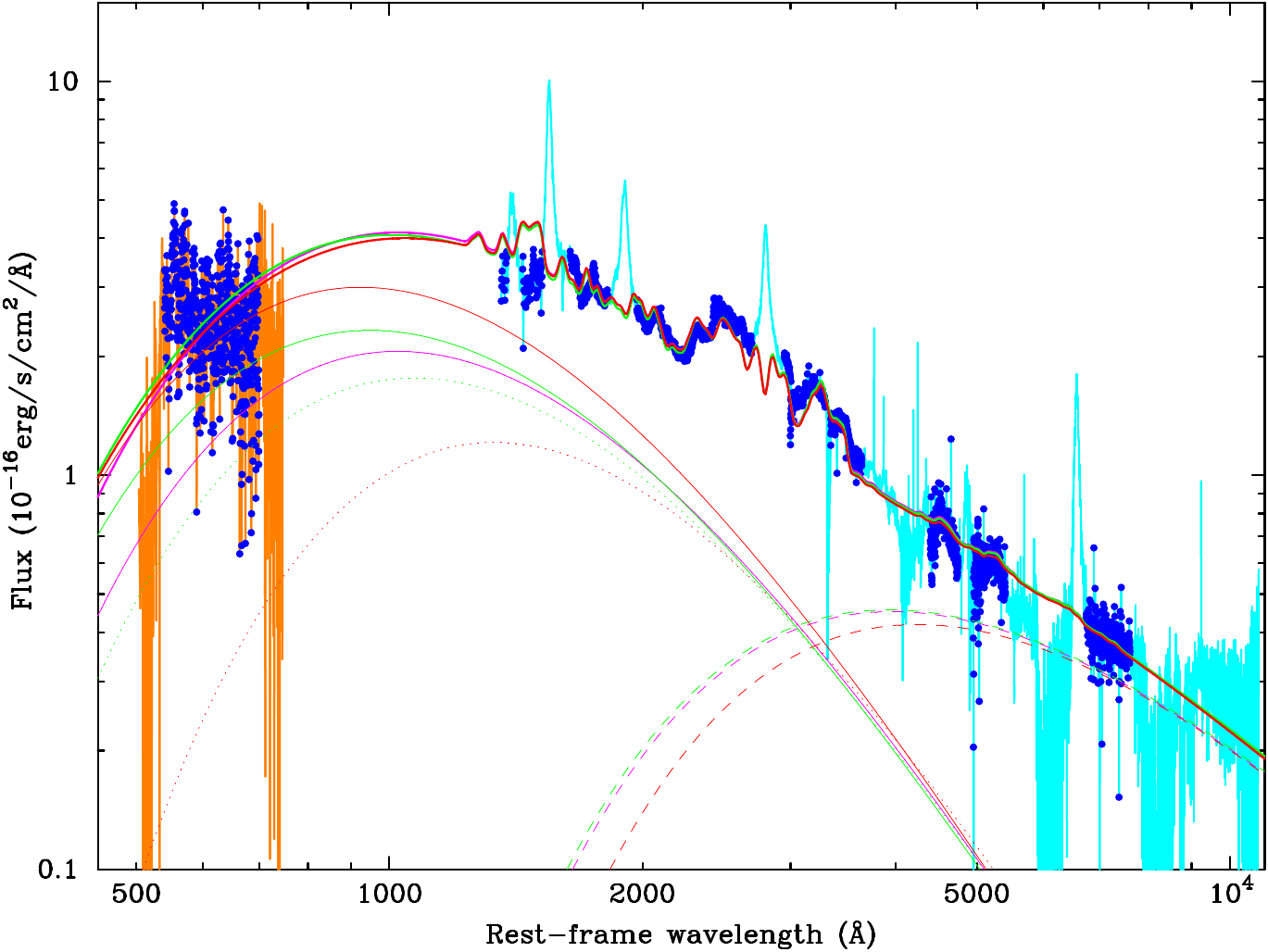}
\caption{
Comparison of the Q J0158-4325 spectrum with the SMBBH-triple disk model. The pink, green, and red lines correspond to models Ma, Mb, and Mc listed in Table \ref{tab:t2}, respectively. Thin solid, dotted, and dashed lines denote emissions from the secondary disk, primary disk, and circumbinary disk, respectively. All other legends are similar to those for Fig.~\ref{fig:SEDall}. 
}
\label{fig:SEDbest}
\end{figure}

\begin{figure*}
\centering
\includegraphics[width=0.35\textwidth,angle=-90]{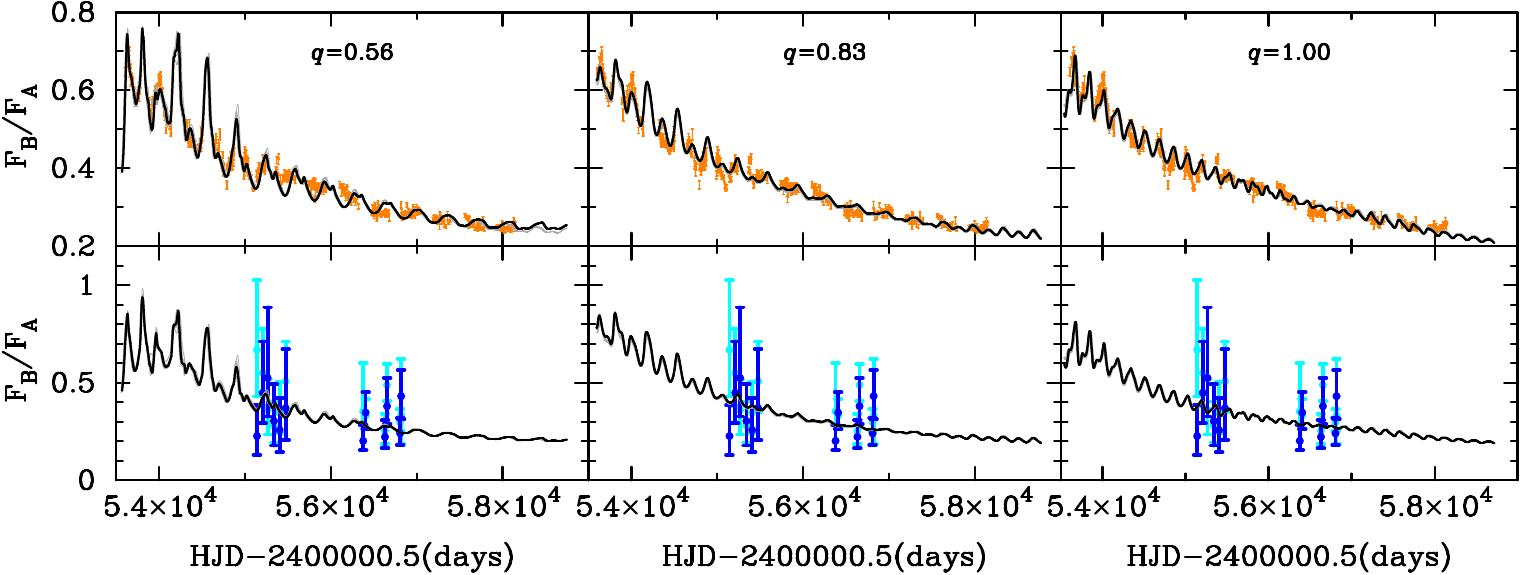}
\caption{
Flux ratio $F_B/F_A$ observed by the Euler telescope from 2005 to 2018 (orange dots with error bars). Blue (cyan) dots with errors are the soft (hard) X-ray flux ratio $F_B/F_A$ as observed by Chandra over the period 2009-2014. The black and gray solid lines represents the corresponding microlensing light curve prediction in the optical and the X-ray band of SMBBHs with the best fitted parameters in Table.\ref{tab:t2}. }
\label{fig:bestlc}
\end{figure*}

From the spectrum optimal fit models, we selected three with varying mass ratios, each exhibiting a reduced chi-squared value of approximately 1.6. The model parameters are shown in Table. \ref{tab:t2}, the predicted theory spectrum are plotted in Figure~\ref{fig:SEDbest} and the optimal fitted microlensing light curve are shown in Figure~\ref{fig:bestlc}. The Ma ($q=1$) case can reproduce the observed light curve very well, in contrast, the Mb and Mc case, despite achieving a comparable chi-square value, produces a theoretical light curve characterized by pronounced alternating peaks: a high peak followed by a slightly lower peak, which deviates from the observational data. So we can say that at least an SMBBH with equal mass can simultaneously match the observed spectrum and microlensing light curve.

\begin{table*}
\centering
\caption{Parameters for the optimal fit cases. 
Here $\tau_{\rm c}$ and $\dot{P}_{\rm orb}$ are the time to coalescence and the changing rate of the orbit period in the observer's frame. The symbol $\chi_{\rm s}^2$ denotes the chi-square value associated with spectrum fitting, while $\chi_{\rm l}^2$ indicates the minimum chi-square value for light curve fitting.}
\begin{tabular}{ccccccccccc} \hline\hline
Model & $M_{\bullet}(\msun)$ & $q$ & $f_{\rm E,1}$ & $f_{\rm E,2}$ & $f_{\rm E,c}$ & $\abbh(r_{\rm g})$&$\tau_{\rm c}$(yr)&$\dot{P}_{\rm orb}$(days/orbit)&$\chi_{\rm s}^2/N_{\rm dof}$&$\chi^2_{\rm l}/N_{\rm dof}$\\ \hline
Ma & $2.79\times10^9$ & 1.0&0.020   & 0.020  & 0.020 &28  &50&-2.4&1.61&8.0\\
Mb & $2.68\times10^9$ & 0.83 &0.018  & 0.026  & 0.022  &29 & 54 &-2.3&1.58&8.1\\
Mc & $3.51\times10^9$ & 0.56&0.008   & 0.025  & 0.015  &24  &37&-3.3&1.59&13.1\\ \hline\hline
\end{tabular}
\label{tab:t2}
\end{table*}

\subsection{X-ray emission and micolensing light curves}

Microlensing can also cause fluctuations in the X-ray band, thus it is also interesting to investigate the X-ray microlensing light curve for Q J0158-4325. We consider the X-ray emission to be uniform and concentrated around the black hole, with a characteristic size of $20GM_{\bullet,i}/c^2$, where $i$ refers to the specific black hole component under consideration. Such a setting is based on studies by \citet[][see also \citealt{,Dai10,MacLeod15}]{Morgan08}, in which the region responsible for X-ray emission is typically constrained to less than $20r_{\rm g}$. We infer X-ray luminosities by applying the empirical relationship between X-ray luminosity and UV luminosity for quasars, using the UV luminosities derived from our accretion model. For this purpose, we apply the relation $\log L_{\nu}(2 {\rm keV}) = 0.85\log L_{\nu}(2500\mathring{A}) + 0.985$ given by \cite{Gupta24}. We calculate $L_{\nu}(2500\mathring{A})$ for each mini-disk and subsequently determine their corresponding X-ray luminosities. By dividing these luminosities by their respective areas, we can derive the surface brightness for both mini-disks. We then convolve the X-ray surface brightness map with the magnification map to obtain the X-ray microlensing light curves. The X-ray light curves for Models M1 to M8 are illustrated in Figure \ref{fig:allmass}, while those for Models Ma to Mc are shown in Figure \ref{fig:bestlc}. Apparently, the microlensing light curves in the X-ray band exhibit the same period and phase as those in the optical band; however, they have a larger amplitude of variation due to their smaller emission region compared to those for the optical wavelengths. 

We collect the X-ray observation data of Q J0158-4325 from \citet[][Table~3 and Fig.~2 therein]{Guerras2017} and compare them with our model. Due to the limitations of observational data we neglect the time delay between the two image as it is significantly smaller than the oscillation period. While this will add some ``noise'' to the light curves. As shown in Figures~\ref{fig:allmass} and \ref{fig:bestlc}, 
although the predicted X‑ray light curve has a larger amplitude than the optical one, the low quality of the X-ray data prevents a reliable determination of its phase or periodicity and thus precludes assessing consistency with the SMBBH scenario.

The X-ray emission of Q J0158-4325 image A in $0.4-1.3$ keV is $0.99\times10^{-13}{\rm erg~s^{-1}cm^{-2}}$ and photo index $\Gamma=1.93$ \citep[see][]{Chen2012}, corresponding to a rest-frame macrolensing-corrected $L_{\nu}(\rm 2keV)\approx4.8\times10^{26}erg~s^{-1}Hz^{-1}$ the macrolensing-corrected $L_{\nu}(2500\AA)=5.1\times10^{30}{\rm erg~s^{-1}Hz^{-1}}$, so the  $\alpha_{\rm ox}=0.3838\log(L_{\nu}(\rm 2keV)/L_{\nu}(2500\AA))\approx-1.5$ which is cosistent with the relation between $\alpha_{\rm ox}$ and monochromatic $2500\AA$ luminosity given by \citet[][Eq.(7) therein]{Gupta24}. 
From the standpoint of X-ray spectrum, the data are internally consistent and show no contradiction with our SMBBH model; however, the current X-ray observations of Q J0158-4325 do not provide additional parameter constraints to the model, if without higher-precision, high cadence X-ray light curves.

\section{Discussion}
\label{sec:dis}

\subsection{The binary orbit period and the semi-major axis}

\begin{figure}
\centering
\includegraphics[width=0.7\textwidth,angle=-90]{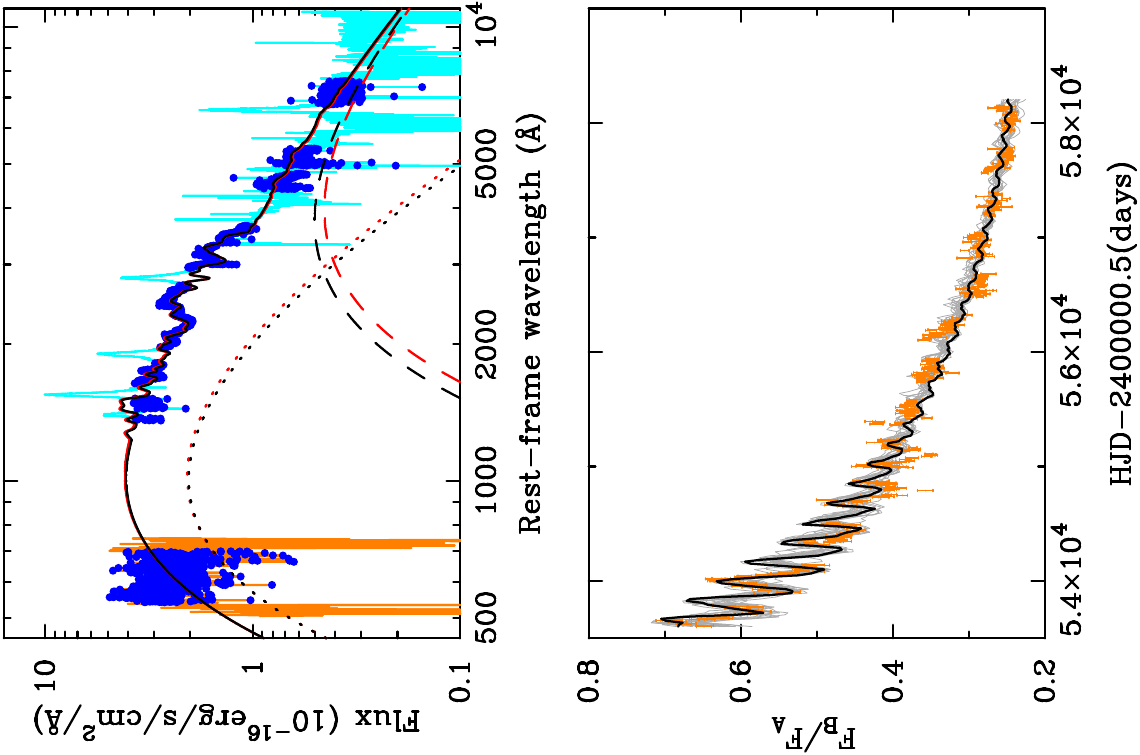}
\caption{
Top Panel: This panel presents a comparison of spectra for Model Ma, as detailed in Table \ref{tab:t2}, with and without accounting for changes in the semi-major axis. The red line indicates the spectrum without consideration of these changes, while the black line represents the spectrum that includes variations in the semi-major axis. The dotted and dashed lines correspond to emissions from the mini disk and the circumbinary disk, respectively. Additional legends are consistent with those in Fig.~\ref{fig:SEDall}. Bottom Panel: This panel illustrates the fitting of the light curve for Model Ma, incorporating the effects of periodic variations. The gray lines indicate the 20 fits with the lowest $\chi^2$, and the black line marks the best fit among them.
}
\label{fig:change}
\end{figure}

The semi-major axis of the binary system Q J0158-4325 is approximately 30 times the gravitational radius, $r_{\rm g}$. Due to this relatively small semi-major axis, the orbital evolution of the system is predominantly affected by the emission of gravitational waves. In the lowest-order post-Newtonian approximation, the rate of change of the semi-major axis resulting from gravitational radiation can be expressed as follows \cite{Peters64}:
\be
\abbh(t)=a_{\rm BBH,0}\left(\frac{\tau_{c,0}-(t-t_0)}{\tau_{c,0}}\right)^{1/4}
\ee
where 
\bea
\tau_{\rm c,0}&=&\frac{5}{256} \frac{c^5 a_{\rm BBH,0}^4}{G^3 \mbh^3}\frac{(1+q)^2}{q} \nonumber \\
&\approx&40.6~ {\rm yr}\frac{(1+q)^2}{4q}\left(\frac{M_{\bullet}}{10^9\msun}\right)^{-5/3} \left(\frac{P_{\rm orb}}{100{\rm day}}\right)^{8/3}
\label{eq:taugw}
\eea
is the gravitational wave-driven merger timescale in the rest frame, and $a_{\rm BBH,0}$ is the semi-marjor axis of SMBBH system at the initial time ($t_0$) in the rest frame. The rate of periodic change can be estimated by
\bea
\frac{\dot{P}_{\rm orb}}{P_{\rm orb}}=-\frac{96}{5}\frac{q}{(1+q)^2}\frac{G^{5/3}\mbh^{5/3}}{c^5}\left(\frac{P_{\rm orb}}{2\pi}\right)^{-8/3}.
\eea
The microlensing light curve of Q J0158-4325, monitored from 2003 to 2018, indicates that variations in the system's semi-major axis could potentially have observable effects. For this nearly equal-mass system, where $P_{\rm orb}=2P_{\rm obs}$, the observer-frame orbital period at 2005 ($t_0$) is approximately 345 days. This period is estimated to be changing at a rate of 2–3 days per orbit, implyinga cumulative change of approximately $10\%$ over the 13-year period. However, the data quality is sufficient to accurately distinguish the orbital period only during the first four years. Over this shorter interval, the period changes by only about $3\%$, a shift too small to significantly affect light curve fits. To explicitly account for this evolution, we adjusted our fitting procedure for model Ma (Fig.~\ref{fig:change}, bottom panel), confirming the robustness of the results.

Additionally, optical spectra obtained in 2019 and UV spectra in 2023 were analyzed to assess any potential spectral impacts. We simulated the spectrum for $t=t_0+14{\rm yr}$ to evaluate these effects. The top panel of Figure~\ref{fig:change} demonstrates that the differences in the spectra, with and without changes in the semi-major axis, are minimal. In summary, despite measurable evolution in the orbital period, its effects on both the microlensing light curve and the spectra over the decade of observation are negligible.

\subsection{The Decoupling of Binaries from Their Disks}

Our analysis rests on the assumption that the binary black holes and the accretion disk remain coupled; that is, the disk evolves on a timescale comparable to the rate of change of the binary's semi-major axis. The validity of this assumption can be assessed through the concept of a decoupling radius $a_{\rm d}$. As shown analytically and numerically by \cite{Dittmann2023}, decoupling occurs when the disk's viscous inflow rate matches the orbital evolution rate of the binary, at a semi-major axis of $a_{\rm d}\approx r_{\rm g}\sqrt{32 f_{\rm in}/15\nu_0}$, where $\nu=\nu_0 GM_{\bullet}/c$ is the kinematic viscosity. Their simulations found that for a low-viscosity case ($\nu_0=0.003$), decoupling occurs at $a_{\rm d}=56r_{\rm g}$, while for a high-viscosity case ($\nu_0=0.03$), the decoupling radius shrinks to
$a_{\rm d}=14r_{\rm g}$. Critically, this high-viscosity decoupling radius is smaller than the inferred semi-major axis of Q J0158-4325. This comparison confirms that our initial assumption of a coupled disk-binary system is reasonable.

\subsection{Radiation model of accretion disk}

Note that our analysis assumes the accretion disk radiation follows the standard disk model.
When the primary black hole’s accretion rate falls below $\sim0.01$ times the Eddington rate, the accretion disk will transition to advection‑dominated accretion flow (ADAF) , which is a kind of geometrically thick,
optically thin, hot accretion flow. ADAF shows very different emission properties compared with thin disk, such as the lower radiative 
efficiency, non-thermal broadband emission etc. \citep{NM2008,Yuan2014}. 
Our fitting shows that the accretion rate is above $0.01$ Eddington rate, so it is reasonable that we always take thin disk in the calculations.

\cite{Laor2014} showed that strong disk winds or outflows can lower the effective temperature of the inner accretion flow, suppressing its UV continuum relative to a standard thin disk. The wind effect becomes negligible for black holes with masses  $\gtrsim 10^9 M_{\odot}$, but for masses near $10^8 M_{\odot}$ the UV luminosity can be reduced by a factor of $2-3$ relative to that from the standard thin disk. Assuming that the SMBBH has a substantially lower mass less than that obtained from our SMBBH model fitting, e.g., $\sim10^9\msun$ as that from Mg II line, with $q\lesssim0.5$ (see model M4 in Figs.~\ref{fig:allmass} and \ref{fig:SED}), then to fit the SED , we may need a suppression factor of the UV emission by disk wind to be $\gtrsim 5$, which seems not likely according to \citet{Laor2014}. Therefore, we conclude our constraints on the SMBBH mass and mass ratio may be not affected much even considering a much complicated accretion disk model with wind and outflows. 

In addition, our microlensing analysis is expected to exhibit only weak sensitive to the detailed form of the disk temperature profile and to the fixed Eddington ratio prior. For our best-fitting configuration, the binary separation is $a_{\rm BBH} \sim 43\,r_{\rm g}$, and the truncation radii of the two mini-disks are $\sim 10^{15}\,\mathrm{cm}$, corresponding to a scale of $r/R_{\rm E} \sim 0.1$. In this regime, modifying the disk model (e.g., by adopting an alternative temperature slope $T(r)\propto r^{-\beta}$ or varying the accretion rate) results primarily in modest changes to the half-light radii of the mini-disks. The half-light radius governs the sharpness of the microlensing magnification peaks within each orbital cycle (i.e., whether the peaks appear slightly sharper or flatter), whereas the presence and period of quasi-periodic modulation are determined by the binary orbital period. Consequently, the overall periodic behaviour of the light curve is dominated by binary dynamics and only mildly affected by reasonable variations in the disk temperature–radius relation or Eddington ratio. Fixing these priors therefore does not qualitatively alter our main conclusions.

\subsection{Alternative scenarios}

An important question is whether the observed periodic signal for Q~J0158$-$4325 could arise from alternative scenarios unrelated to an SMBBH system. \citet{Millon2022} has examined this issue for several apparently alternative scenarios and found them unlikely. We briefly summarize them as follows.

\begin{itemize}
\item
Quasar stochastic variability can produce light curves that mimic periodic behavior. \citet{Millon2022} performed a generalized Lomb–Scargle analysis to mock light curves by modelling quasar variability as a damped random walk with a reverberated broad line region (BLR) component and found that the probability to produce a spectral peak stronger than that in the 2005–2011 light curve is only $0.6\%$ (a $\sim 3.7\sigma$ rejection). In addition, the oscillations occur only during the high-magnification microlensing event and their amplitude scales with the microlensing magnification, behaviour not expected for red-noise intrinsic variability. Therefore, it is unlikely that the observed periodic variation is due to stochastic variation of the quasar.
\item
Microlensing of emission from a single-black-hole disk by a pair of microlenses (e.g., binary star or star-planet binary) would lead to periodic variations. However, \citet{Millon2022} showed that such a binary microlensing can only produces periodic modulation of a magnitude $\lesssim 10^{-5}$ with the observed period or $\sim 0.2$\,mag with a period of $\sim2000$\,yr, incompatible with the $172.6$-day period. These analyses suggest that binary microlenses could not simultaneously reproduce the observed amplitude and period.
\item
Intermediate-mass black holes (IMBHs) embedded in the accretion disk carve gaps and carry their own compact accretion flows, thereby periodically modulating the microlensed surface brightness. The associated emitting regions of such an IMBH is roughly
\[
r_{1/2} \approx 5\times 10^{13} \,\mathrm{cm} \left( \frac{M_\bullet}{10^5 \, M_\odot} \right)^{2/3} \left( \frac{f_{\rm E}}{\epsilon} \right)^{1/3} \left( \frac{\lambda}{\mu \mathrm{m}} \right)^{4/3},
\]
far smaller than the $\sim 10^{15}\,\mathrm{cm}$ mini-disk sizes inferred in our SMBBH model and also much smaller than $R_{\rm E}$. Such compact sources would generate sharp, spiky magnification events when traversing caustics, distinct from the relatively smooth peaks observed in the 2005–2011 data, which are well reproduced by extended mini-disks with half-light radii of order $10^{15}\,\mathrm{cm}$ (see also 
\citealt{Millon2022}). Thus, while embedded IMBHs or other compact perturbers can in principle modulate the disk emission, their characteristic half-light radii are too small to match the observed peak shapes and amplitudes.
\end{itemize}
Taken together, constraints on stochastic intrinsic variability, binary microlenses, and embedded IMBHs indicate that these alternative scenarios cannot plausibly reproduce the observed periodic microlensing signal, supporting a source-plane origin tied to an SMBBH system.

\subsection{Detection probability}

For an SMBBH system with a merger timescale on the order of decades to centuries, the probability of observing a merger is extremely low. Estimates suggest that the typical net lifetime of normal quasars ranges from approximately $10^7$ to $10^8$ years \citep[e.g.,][]{YT02,Martini2004,YL08}. Consequently, the occurrence rate of SMBBHs with merger timescale the same as Q J0158-4325 is estimated to be at most between $10^{-5}$ and $10^{-6}$. With an expected population of $10^5$ strongly lensed AGNs \citep[e.g.,][]{Oguri2010}, we anticipate that up to $0.1-1$ such SMBBHs can be observed via microlensing variations. Consequently, the probability of such a detection is so low that it would be a matter of chance. However, we should note that SMBBH-triple disk systems with much smaller masses $\lesssim 10^8M_\odot$ and larger separation $\sim 100r_{\rm g}$ can have substantially longer merger timescale (see Eqs.~\ref{eq:taugw} and \ref{eq:abbh}) and thus much larger occurrence rate. In this case, it is should be promising to observe them via the microlensing variations like Q J0158-4325 among lensed AGNs .

\section{Conclusions }
\label{sec:con}

In this paper, we present self-consistent evidence that the observed properties of the lensed quasar Q J0158-4325 are explained by an SMBBH system with a triple-disk accretion structure. Our physical model, which incorporates realistic disk geometries, orbital dynamics, and microlensing effects, reproduces the $\sim$173-day periodic variations seen in the long-term optical microlensing light curve. By jointly analyzing the microlensing variability and the optical-UV SED, we substantially reduce key parameter degeneracies and constrain the system's physical properties. Our analysis favors models with a large total SMBBH mass (of order $\sim 10^{9.5}M_\odot$) and a high mass ratio ($q>0.5$), a conclusion supported by Bayesian model comparison. While optical data alone permit both high- and low-mass ratio models, incorporating HST UV data effectively disfavors low mass ratio scenarios ($q<0.5$), which overpredict UV flux. The Bayes factor distinctly favors high mass ratio models, highlighting the critical role of multi-wavelength data—especially in the UV—for accurately constraining accretion disk structure in these complex systems.

Additionally, our model generates testable predictions. The expected microlensing signals in the X-ray band should match the same period and phase as the optical variations but with a larger amplitude due to the more compact emission region. This highlights the need for future high-cadence X-ray monitoring, combined with optical observations, to detect and characterize SMBBH systems through microlensing-induced variability, although current X-ray data are too limited to confirm the expected microlensing signatures in Q J0158-4325.

Our analysis of Q J0158-4325 highlights the effective combination of microlensing and spectral modeling for understanding the environment around SMBBHs. It not only offers a coherent model for this specific system but also establishes a broader methodology for identifying and studying SMBBHs in lensed quasars. 
This approach is particularly timely given upcoming time-domain surveys such as the Rubin Observatory Legacy Survey of Space and Time (LSST), which will monitor thousands of lensed quasars across six optical/UV bands over a ten-year baseline and thus provide a rich dataset to enable systematic searches for periodic signals and microlensing-induced variability linked to SMBBHs.

The most likely models for Q J0158–4325 suggest a gravitational-wave merger time of just a few decades. Although the gravitational-wave frequency of this system still lies outside the sensitivity band of LISA, analogous systems with significantly lower masses at later evolutionary stages could be promising targets for future low-frequency gravitational-wave detectors. More broadly, the electromagnetic identification of SMBBHs in lensed quasars, as demonstrated in this work, will provide valuable targets and information to support multi-messenger studies with missions such as LISA.
However, the low probability of observing a binary during such a brief evolutionary phase raises the question of whether this detection was simply a matter of extraordinary luck. This question, together with the imminent advent of LSST and future gravitational-wave facilities, motivates systematic searches using our framework to determine how frequently such imminent mergers occur and to build a statistically meaningful sample of SMBBH candidates.

\section*{Acknowledgements}

This work is partly supported by the National Key Research and Development Program of China (grant nos. 2020YFC2201400, 2022YFC2205201, and 2023YFA1607904), the National SKA Program of China (grant no. 2020SKA0120101), the National Natural Science Foundation of China (grant nos. 12273050), the Strategic Priority Research Program of the Chinese Academy of Sciences (grant no. XDB0550302), the National Astronomical Observatory of China (grant no. E4TG660101, E4ZR0510), the Beijing Municipal Natural Science Foundation (grant no. 1242032), the Youth Innovation Promotion Association of the Chinese Academy of Sciences (No. 2022056), and the China Manned Space Program with grant no. CMS-CSST-2025-A07. This work was performed in part at the Aspen Center for Physics, which is supported by National Science Foundation grant PHY-2210452.

\section*{Data Availability}

The data of optical specturm underlying this article are available in ESO Archive Science Portal at 
https://archive.eso.org/scienceportal/home, and can be accessed with object name QJ0158-4325. 

The data of UV specturm underlying this article are available in MAST: Barbara A. Mikulski Archive for Space Telescopes at https://mast.stsci.edu/portal/Mashup/Clients/Mast/Portal.html, and can be accessed with object name QSO J0158-4325.



\bibliographystyle{mnras}
\bibliography{bib} 







\bsp	
\label{lastpage}
\end{document}